\documentclass[%showpacs, showkeys,
12pt,
preprint,preprintnumbers,nofootinbib,
groupedaddress,superscriptaddress,amsmath,amssymb]{revtex4}
\usepackage{graphicx}% Include figure files
\usepackage{dcolumn}% Align table columns on decimal point
\usepackage{bm}% bold math
\usepackage{amssymb}
\usepackage{amsmath}
\usepackage{epsfig}    
\usepackage{color}
\usepackage{hhline}
\def\be{\begin{equation}}
\def\ee{\end{equation}}
\newcommand{\bea}{\begin{eqnarray}}
\newcommand{\eea}{\end{eqnarray}}
\newcommand{\nn}{\nonumber}

\numberwithin{equation}{section}

\begin{document}
 {\begin{flushright}{KIAS-P20035, APCTP Pre2020 - 014}\end{flushright}}
%%%%%%%%%
\title{A linear seesaw model with $A_4$-modular flavor and local $U(1)_{B-L}$ symmetries}
% A modular $A_4$ symmetric scotogenic model}
%

\author{Takaaki Nomura}
\email{nomura@kias.re.kr}
\affiliation{School of Physics, KIAS, Seoul 02455, Republic of Korea}

\author{Hiroshi Okada}
\email{hiroshi.okada@apctp.org}
\affiliation{Asia Pacific Center for Theoretical Physics (APCTP) - Headquarters San 31, Hyoja-dong,
Nam-gu, Pohang 790-784, Korea}
\affiliation{Department of Physics, Pohang University of Science and Technology, Pohang 37673, Republic of Korea}

\date{\today}

\begin{abstract}
{
We discuss a linear seesaw model with local $U(1)_{B-L}$ and modular $A_4$ symmetries.
The neutrino mass matrix for linear seesaw mechanism is realized by $U(1)_{B-L}$ charge assignment and the nature of modular $A_4$ symmetry.
We formulate neutrino mass and carry out numerical analysis showing some predictions for observables in neutrino sector.
%%%
{\it Remarkably, the case of inverted neutrino mass ordering (IO) is realized by a specific region at nearby $\tau=\omega=e^{2\pi i/3}$, which is favored by a string theory. Thus, our prediction would be very strong in case of IO.}
}
\end{abstract}
\maketitle
\newpage
\section{Introduction}\label{sec1}

The understanding of flavor structure is one of the important issues in particle physics since we do not have any symmetry to control flavor in the standard model (SM).
Thus introduction of a flavor symmetry is typical strategy in constructing a model of physics beyond the SM.

One of the interesting approach is application of modular flavor symmetries proposed by~\cite{Feruglio:2017spp, deAdelhartToorop:2011re} to describe flavor structures. 
In this framework, a coupling can be transformed under a non-trivial representation of a non-Abelian discrete group and we can realize flavor structure without many scalar fields such as flavons.  
Then some typical groups are found to be available in basis of the modular group $A_4$~\cite{Feruglio:2017spp, Criado:2018thu, Kobayashi:2018scp, Okada:2018yrn, Nomura:2019jxj, Okada:2019uoy, deAnda:2018ecu, Novichkov:2018yse, Nomura:2019yft, Okada:2019mjf,Ding:2019zxk,Nomura:2019lnr,Kobayashi:2019xvz,Asaka:2019vev,Zhang:2019ngf, Gui-JunDing:2019wap,Kobayashi:2019gtp,Nomura:2019xsb,Behera:2020sfe,Wang:2019xbo,Okada:2020dmb,Okada:2020rjb}, $S_3$ \cite{Kobayashi:2018vbk, Kobayashi:2018wkl, Kobayashi:2019rzp, Okada:2019xqk}, $S_4$ \cite{Penedo:2018nmg, Novichkov:2018ovf, Kobayashi:2019mna,King:2019vhv,Okada:2019lzv,Criado:2019tzk,Wang:2019ovr}, $A_5$ \cite{Novichkov:2018nkm, Ding:2019xna,Criado:2019tzk}, larger groups~\cite{Baur:2019kwi}, multiple modular symmetries~\cite{deMedeirosVarzielas:2019cyj}, and double covering of $A_4$~\cite{Liu:2019khw} and $S_4$~\cite{Novichkov:2020eep, Liu:2020akv} in which  masses, mixing, and CP phases for quark and/or lepton are predicted.~\footnote{Some reviews are useful to understand the non-Abelian group and its applications to flavor structure~\cite{Altarelli:2010gt, Ishimori:2010au, Ishimori:2012zz, Hernandez:2012ra, King:2013eh, King:2014nza, King:2017guk, Petcov:2017ggy}.}
Furthermore, a systematic approach to understand the origin of CP transformations has been discussed in ref.~\cite{Baur:2019iai}, 
and CP violation in models with modular symmetry is also discussed in Ref.~\cite{Kobayashi:2019uyt,Novichkov:2019sqv}, 
and a possible correction from K\"ahler potential is also discussed in Ref.~\cite{Chen:2019ewa}.
In particular, it is interesting to apply a modular symmetry in constructing a new physics model for neutrino mass generation in which 
we would obtain prediction for signals of new physics correlated with observables in neutrino sector.

In this study, we construct a linear seesaw model with local $U(1)_{B-L}$ and modular $A_4$ symmetry~\footnote{A linear seesaw model with modular $A_4$ and global symmetry is found in ref.~\cite{Behera:2020sfe} providing different flavor structure of neutrino mass and predictions from ours. }.
In our scenario desired mass matrix for linear seesaw mechanism~\cite{Wyler:1982dd,Akhmedov:1995ip,Akhmedov:1995vm} can be realized by $U(1)_{B-L}$ charge assignment and the nature of modular $A_4$ symmetry.
We then formulate neutrino mass matrix under the symmetry and carry out numerical analysis searching for parameters fitting neutrino measurements.
Our numerical analysis shows some predictions for observables in neutrino sector.

This paper is organized as follows. In Sec.~\ref{sec:model} we introduce our model and formulate neutrino mass from linear seesaw mechanism with modular $A_4$ symmetry.
 In Sec.~\ref{sec:results} we carry out numerical analysis and show correlations between observables in the neutrino sector, and conclude our results in Sec.~\ref{sec:conclusion}.

\section{Model}
\label{sec:model}
%
% \begin{widetext}
\begin{center} 
\begin{table}[t!]%[tbc]
%\begin{tiny}
\begin{tabular}{|c||c|c|c|c|c|c|c||c|c|c|c|}\hline\hline  
  & \multicolumn{7}{c||}{Fermions} & \multicolumn{3}{c|}{Scalars} \\ \hline \hline
& ~$Q_L$~& ~$u_R$~  & ~$d_R$~& ~$L_L$~& ~$[e_R^c,\mu_R^c,\tau_R^c]$~& ~$[N_{R_1}^c,N_{R_2}^c,N_{R_3}^c]$~& ~$[S_{L_1},S_{L_2},S_{L_3}]$~& ~$H_1$~& ~$H_2$~& ~$\varphi$~ \\ \hline \hline 
%%%
$SU(3)_C$ & $\bm{3}$  & $\bm{3}$ & $\bm{3}$ & $\bm{1}$ & $\bm{1}$ & $\bm{1}$ & $\bm{1}$ & $\bm{1}$& $\bm{1}$ & $\bm{1}$   \\\hline 
$SU(2)_L$ & $\bm{2}$  & $\bm{1}$  & $\bm{1}$  & $\bm{2}$  & $\bm{1}$  & $\bm{1}$  & $\bm{1}$ & $\bm{2}$   & $\bm{1}$& $\bm{1}$     \\\hline 
$U(1)_Y$   & $\frac16$ & $\frac23$ & $-\frac13$ & $-\frac12$  & $1$ & $0$  & $0$  & $\frac12$ & $\frac12$& $0$  \\\hline
$U(1)_{B-L}$   & $\frac13$ & $\frac13$ & $\frac13$   & $-1$  & $1$  & $1$  & $0$ & $0$  & $1$ & $-1$   \\\hline
$A_4$ & $\bm{1}$ & $\bm{1}$ & $\bm{1}$ & ${\bf 3}$ & ${\bf 1}, {\bf 1^{\prime\prime}}, {\bf 1^{\prime}}$ & $\bm{3}$ & ${\bf 1}, {\bf 1^{\prime}}, {\bf 1^{\prime\prime}}$ & $\bm{1}$ & $\bm{1}$ \ & $\bm{1}$ \\ \hline
$-k_I$ & $0$ & $0$ & $0$ & $-1$ & $-1$ & $-1$ & ${-1}$ & $0$ & $0$ & $0$ \\
\hline
\end{tabular}
\caption{Particle content of the Standard Model extended with two types of sterile neutrinos $N_R, S_L$ and extra singlet scalar $\varphi$ for implementation of inverse seesaw mechanism and their charge assignments under $SU(3)_C\times SU(2)_L\times U(1)_Y\times U(1)_{B-L}\times  A_4\times k_I$ where $k_I$ is the number of modular weight.}
\label{tab:fields-inverse}
% \end{tiny}
\end{table}
\end{center}
%\end{widetext}
%
In this section we briefly discuss the model framework for linear seesaw mechanism introducing $B-L$ local Abelian symmetry $U(1)_{B-L}$ and modular $A_4$ symmetry. 
In the model, we introduce three families of right(left)-handed $SU(2)$ singlet fermions $N_R(S_L)$ with $-1$(0) charge under the $U(1)_{B-L}$ gauge symmetry,
and an isospin singlet fields $\varphi$ with $-1$ charge under the same $U(1)$ symmetry.
 Furthermore, two Higgs doublet $H_1$ and $H_2$ are introduced where $H_2$ also has charge 1 under $U(1)_{B-L}$ while $H_1$ has no $B-L$ charge to induce the masses of SM fermions from the Yukawa Lagrangian after the spontaneous symmetry breaking as in the SM. 
We also assign modular weight $k_I$ to these fields as summarized in Table~\ref{tab:fields-inverse}.
Also only lepton doublet $L_L$ and right-handed sterile neutrino $N_R^c$ are chosen to be $A_4$ triplet.
Then we impose Yukawa Lagrangian should be invariant under these symmetries where each term has vanishing modular weight; properties of modular symmetry are referred to Appendix.  
 Here we denote each of vacuum expectation value (VEV) to be $\langle H_{1,2} \rangle\equiv [0,v_{1,2}/\sqrt2]^T$, and $\langle \varphi \rangle\equiv v_{\varphi} /\sqrt2$.
%%%
The SM singlet scalar $\varphi$ plays a role in inducing $(H_1^\dagger H_2) \varphi$ term in order to avoid massless CP-odd scalar from Higgs doublets. 
The mass scale of the SM singlet scalars are taken to be much higher than electroweak scale and we obtain well-known two Higgs doublet potential after they develop VEVs.
Also $Z'$ boson from $U(1)_{B-L}$ gets mass by singlet scalar VEV, and we just assume the mass and gauge coupling satisfy current experimental constraints.
In this paper, we omit the details of the scalar/gauge sector and focus on the neutrino sector.

Using the particle contents and symmetries mentioned in Table \ref{tab:fields-inverse},  
the relevant Yukawa Lagrangian for leptons--including charged leptons and neutral leptons-- can be written as, 
\begin{align}\label{a4lag}
 -\mathcal{L}_{\rm lepton} = \mathcal{L}_{M_\ell} +\mathcal{L}_{M} +\mathcal{L}_{\rm M'_D} + \mathcal{L}_{\rm M_D},
\end{align}
where $\mathcal{L}_{M_\ell} $ is Yukawa Lagrangian inducing charged lepton masses, $ \mathcal{L}_{\rm 
M_D}$ is for Dirac neutrino mass term connecting active light neutrinos $\nu_L$ and $N_R$, $\mathcal{L}_{\rm M}$ is for mixing term between two types of sterile neutrinos $N_R$ and $S_L$, 
and $\mathcal{L}_{\rm M'_{D}}$ is for mass term connecting $\nu_L$ and $S_L$. 
The Majorana mass terms for the sterile neutrinos $N_R$ and $S_L$ are absent; 
the former one is forbidden by $U(1)_{B-L}$ symmetry 
and the latter one cannot be constructed due to the nature of modular $A_4$ symmetry since 
$A_4$ singlets have to have 4 modular weights at least.
%modular form with weight 2 is only $A_4$ triplet. 

\noindent \\
{\bf \underline{Charged lepton mass matrix}}: \\
In this model lepton doublets constitute $A_4$ triplet as $L_L\equiv [L_{L_e},L_{L_\mu},L_{L_\tau}]^T$ and $\bar L_L\equiv [\bar L_{L_e},\bar L_{L_\tau},\bar L_{L_\mu}]^T$. 
Similar to $L_L$, modular couplings are also defined by $Y^{(2)}_{\bf 3}\equiv [y_1,y_2,y_3]^T$ and $Y^{(2)*}_{\bf 3}\equiv [y_1^*,y_3^*,y_2^*]^T$ under $A_4$ triplet. 
The Lagrangian to give the charged-lepton mass matrix is given by 
\[\mathcal{L}_{M_\ell} =Y^{(2)*}_{\bf3}\otimes \bar L_L\otimes e_R\otimes H_1\]
 that is explicitly written in terms of three free parameters, requiring invariance under $SU(3)_C\times SU(2)_L\times U(1)_Y\times U(1)_{B-L}\times  A_4\times k_I$, as follows:
\begin{align}
& a_\ell(y^*_1 \bar L_{L_e} +y^*_2 \bar L_{L_\tau} +y^*_3 \bar L_{L_\mu}) e_{R} H_1 + b_\ell(y^*_3 \bar L_{L_\tau} +y^*_1 \bar L_{L_\mu} +y^*_2 \bar L_{L_e}) \mu_{R} H_1 \nn \\
&+ c_\ell(y^*_2 \bar L_{L_\mu} +y^*_1 \bar L_{L_\tau} +y^*_3 \bar L_{L_e}) \tau_{R} H_1.
\end{align}
Then the mass matrix for charged-lepton in basis of $[e,\mu,\tau]$ is given by
\begin{align}
(M_\ell)_{LR} =\frac{v_1}{\sqrt2}
 \begin{pmatrix}  y^*_1  &  y^*_2 &  y^*_3 \\
y^*_3  &  y^*_1 &  y^*_2 \\
y^*_2  &  y^*_3  & y^*_1
\end{pmatrix} 
\begin{pmatrix}  a_\ell  &  0 &  0 \\
0  &   b_\ell   &  0 \\
0  &  0  &   c_\ell \end{pmatrix}                   
\label{Eq:Mell} .
\end{align}
The charged-lepton mass eigenstates are found by diagonalizing ${\rm diag}[m_e,m_\mu,m_\tau] = V_{L_\ell}^\dag M_\ell V_{R_\ell}$, where $V_{L_\ell, R_\ell}$ are unitary matrices. 
%%%
In our numerical analysis, we will determine the free parameters $a_\ell,b_\ell,c_\ell$ so as to fit the three charged-lepton mass eigenstates after giving all the numerical values, by applying the relations:
\begin{align}
&{\rm Tr}[M_\ell {M_\ell}^\dag] = |m_e|^2 + |m_\mu|^2 + |m_\tau|^2,\quad
 {\rm Det}[M_\ell {M_\ell}^\dag] = |m_e|^2  |m_\mu|^2  |m_\tau|^2,\nn\\
&({\rm Tr}[M_\ell {M_\ell}^\dag])^2 -{\rm Tr}[(M_\ell {M_\ell}^\dag)^2] =2( |m_e|^2  |m_\mu|^2 + |m_\mu|^2  |m_\tau|^2+ |m_e|^2  |m_\tau|^2 ).\label{eq:l-cond}
\end{align}

\noindent
{\bf \underline{Neutral fermion mass matrix}:}\\
The Lagrangian to give the neutral mass matrices are given by 
\[ \mathcal{L}_{\rm M_D}+\mathcal{L}_{\rm M'_D} +\mathcal{L}_{M}
= Y^{(2)*}_{\bf3}\otimes \bar L_L\otimes N_R\otimes \tilde H_1
+ 
Y^{(2)*}_{\bf3}\otimes \bar L_L\otimes S_L^c\otimes \tilde H_2 
+ 
Y^{(2)}_{\bf3}\otimes \bar N_R\otimes S_L \otimes \varphi. \]

The first term of the Lagrangian is explicitly written in terms of two parameters that is given by
\begin{align}
\mathcal{L}_{\rm M_D} = &  \biggl[  \frac{\alpha_1}{3}
\left[y^*_1(2\bar L_{L_e} N_{R_1}-\bar L_{L_\mu} N_{R_3}-\bar L_{L_\tau} N_{R_2})  
+
y^*_2 (2\bar L_{L_\mu} N_{R_2}-\bar L_{L_e} N_{R_3}-\bar L_{L_\tau} N_{R_1})\right.\nn\\
&\left. \qquad +
y^*_3 (2\bar L_{L_\tau} N_{R_3}-\bar L_{L_e} N_{R_2}-\bar L_{L_\mu} N_{R_1})
\right]   \nn \\
%\nn\\+&
&+\frac{\alpha_2}{2}
\left[y^*_1(-\bar L_{L_\mu} N_{R_3}+\bar L_{L_\tau} N_{R_2})  
+
y^*_2 (\bar L_{L_e} N_{R_3}-\bar L_{L_\tau} N_{R_1})
%\right.\nn\\&\left.+
+y^*_3 ( \bar L_{L_\mu} N_{R_1}-\bar L_{L_e} N_{R_2})
\right] \biggr] \tilde H_1.
\end{align}
%%%
After the spontaneous symmetry breaking, we obtain the mass matrix
\begin{align}
(m_D)_{\nu_L N_R}  = \frac{v_1}{\sqrt2} 
\left[
\frac{\alpha_1}{3} 
\begin{pmatrix}  
2y^*_1  &  -y^*_3 &  -y^*_2 \\
-y^*_3  &  2y^*_2 &  -y^*_1 \\
-y^*_2  & - y^*_1  & 2y^*_3
\end{pmatrix}  
%%%               
+
\frac{\alpha_2}{2} 
\begin{pmatrix}  
0  &  -y^*_3 &  y^*_2 \\
y^*_3  &  0 &  -y^*_1 \\
-y^*_2  &  y^*_1  &0
\end{pmatrix}  
\right]
\equiv 
\frac{v_1}{\sqrt2}
(\tilde m_D)_{\nu_L N_R} 
\label{Eq:md} .
\end{align}

The second term of the Lagrangian is written in terms of two parameters that is given by
\begin{align}
\mathcal{L}_{\rm M'_D} = &
\beta_1\left[y^*_1 \bar L_{L_e} + y^*_2 \bar L_{L_\tau} + y^*_3  \bar L_{L_\mu}\right] \tilde H_2 S^c_{L_1}
+
\beta_2\left[y^*_2 \bar L_{L_\mu} + y^*_1 \bar L_{L_\tau} + y^*_3  \bar L_{L_e}\right] \tilde H_2 S^c_{L_2}\nn\\
&+ 
\beta_3\left[y^*_3 \bar L_{L_\tau} + y^*_1 \bar L_{L_\mu} + y^*_2  \bar L_{L_e}\right] \tilde H_2 S^c_{L_3}.
\end{align}
%%%
After the spontaneous symmetry breaking, we obtain the mass matrix
\begin{align}
(m'_D)_{\nu_L S_L^c} =\frac{v_2}{\sqrt2}
%%%
\begin{pmatrix}  
y^*_1  &  y^*_3&  y^*_2  \\
y^*_3  &  y^*_2  &  y^*_1\\
y^*_2  &  y^*_1&  y^*_3
\end{pmatrix}  
%%%               
\begin{pmatrix}  
\beta_1  & 0   & 0  \\
0  & \beta_2    & 0 \\
0  & 0    & \beta_3 \\
\end{pmatrix}  
=
\frac{v_2}{\sqrt2}
(\tilde m'_D)_{\nu_L S_L^c}
\label{Eq:mdp} .
\end{align}

The third term of the Lagrangian is written in terms of two parameters that is given by
\begin{align}
\mathcal{L}_{M} = &
\gamma_1\left[y_1 \bar N_{R_e} + y_2 \bar N_{R_\tau} + y_3  \bar N_{R_\mu}\right] S_{L_1}\varphi
+
\gamma_2\left[y_2 \bar N_{R_\mu} + y_1 \bar N_{R_\tau} + y_3  \bar N_{R_e}\right] S_{L_2}\varphi\nn\\
&+
\gamma_3\left[y_3 \bar N_{R_\tau} + y_1 \bar N_{R_\mu} + y_2  \bar N_{R_\tau}\right] S_{L_3}\varphi.
\end{align}
%%%
After the spontaneous symmetry breaking we obtain the mass matrix
\begin{align}
(M)_{N_R S_L} =\frac{v_\varphi}{\sqrt2}
%%%
\begin{pmatrix}  
y_1  &  y_3&  y_2  \\
y_3  &  y_2  &  y_1\\
y_2  &  y_1&  y_3
\end{pmatrix}  
%%%               
\begin{pmatrix}  
\gamma_1  & 0& 0  \\
0  & \gamma_2  & 0\\
0  & 0 & \gamma_3  \\
\end{pmatrix}  
=
\frac{v_\varphi}{\sqrt2} (\tilde M)_{N_R S_L}
\label{Eq:Mell} .
\end{align}

In basis of $[\nu^c_L,N_R,S_L^c]^T$, the neutral fermion mass matrix is given by
\begin{align}
M_N=
 \begin{pmatrix}  0  &  m_D &  m'_D \\
m_D^T  & 0 & M^* \\
m'^T_D  &  M^\dag  & 0
\end{pmatrix}   
=
\frac{v_1v_2}{\sqrt{2} v_\varphi}
 \begin{pmatrix}  
 0  &  \tilde m_D &\tilde m'_D \\
\tilde m_D^T  & 0 & \tilde M^* \\
\tilde m'^T_D  & \tilde M^\dag  & 0
\end{pmatrix}                
\label{Eq:neut} .
\end{align}
Then, block diagonalizing the above matrix, the active neutrino mass matrix is given by
\begin{align}
m_\nu&
= m'_D (M^*)^{-1} m^T_D + [ m'_D (M^*)^{-1} m^T_D]^T =
\frac{v_1v_2}{\sqrt{2} v_\varphi}
\left(
\tilde m'_D (\tilde M^*)^{-1}\tilde m^T_D + [\tilde m'_D (\tilde M^*)^{-1}\tilde m^T_D]^T
\right)\nn\\
%%%
&=%\frac{v_1v_2}{\sqrt{2} v_\varphi}
\kappa \tilde m_\nu, 
\label{Eq:act-neut} 
\end{align}
where $\kappa\equiv \frac{v_1v_2}{\sqrt{2} v_\varphi}$ and we have assumed hierarchy for scale of mass matrices as $m_D, m'_D \ll M$.
Note that such hierarchy of mass matrix can be realized by choosing $v_{1,2} \ll v_\varphi$~\footnote{In some models, hierarchy of mass matrices is realized dynamically~\cite{Wang:2015saa, Das:2017ski}.}.
The neutrino mass eigenstate is found by diagonalizing the mass matrix, $D_\nu=\kappa D_\nu= U_\nu^T m_\nu U_\nu=\kappa U_\nu^T \tilde m_\nu U_\nu$, where $U_\nu$ is a unitary matrix. Then, the Pontecorvo-Maki-Nakagawa-Sakata (PMNS) matrix is given by $U_{PMNS}\equiv V^\dag_{L_\ell} U_\nu$.
% since the charged-lepton is not diagonal basis in original Lagrangian.
Then $\kappa$ is determined by
\begin{align}
{\rm (NO)}:\  \kappa^2= \frac{|\Delta m_{\rm atm}^2|}{\tilde D_{\nu_3}^2-\tilde D_{\nu_1}^2},
\quad
{\rm (IO)}:\  \kappa^2= \frac{|\Delta m_{\rm atm}^2|}{\tilde D_{\nu_2}^2-\tilde D_{\nu_3}^2},
 \end{align}
where $\Delta m_{\rm atm}^2$ is atmospheric neutrino mass difference squared and NO and IO stand for normal and inverted ordering respectively. 
Subsequently, the solar mass difference squared can be written in terms of $\kappa$ as follows:
\begin{align}
\Delta m_{\rm sol}^2= {\kappa^2}({\tilde D_{\nu_2}^2-\tilde D_{\nu_1}^2}),
 \end{align}
 which can be compared to the observed value.
 % 
%Tr$[D_{\nu}] \lesssim$ 0.12 eV is given by the recent cosmological data~\cite{Aghanim:2018eyx, Vagnozzi:2017ovm}.
{ In our model, PMNS matrix is parametrized by three mixing angle $\theta_{ij} (i,j=1,2,3; i < j)$, one CP violating Dirac phase $\delta_{CP}$,
and two Majorana phases $\{\alpha_{21}, \alpha_{32}\}$ as follows:
\begin{equation}
U_{PMNS} = 
\begin{pmatrix} c_{12} c_{13} & s_{12} c_{13} & s_{13} e^{-i \delta_{CP}} \\ 
-s_{12} c_{23} - c_{12} s_{23} s_{13} e^{i \delta_{CP}} & c_{12} c_{23} - s_{12} s_{23} s_{13} e^{i \delta_{CP}} & s_{23} c_{13} \\
s_{12} s_{23} - c_{12} c_{23} s_{13} e^{i \delta_{CP}} & -c_{12} s_{23} - s_{12} c_{23} s_{13} e^{i \delta_{CP}} & c_{23} c_{13} 
\end{pmatrix}
\begin{pmatrix} 1 & 0 & 0 \\ 0 & e^{i \frac{\alpha_{21}}{2}} & 0 \\ 0 & 0 & e^{i \frac{\alpha_{31}}{2}} \end{pmatrix},
\end{equation}
where $c_{ij}$ and $s_{ij}$ stand for $\cos \theta_{ij}$ and $\sin \theta_{ij}$ respectively. 
Then, these mixings are given in terms of the components of $U_{PMNS}$ as follows:
\begin{align}
\sin^2\theta_{13}=|(U_{PMNS})_{13}|^2,\quad 
\sin^2\theta_{23}=\frac{|(U_{PMNS})_{23}|^2}{1-|(U_{PMNS})_{13}|^2},\quad 
\sin^2\theta_{12}=\frac{|(U_{PMNS})_{12}|^2}{1-|(U_{PMNS})_{13}|^2}.
\end{align}
Also we compute the Jarlskog invariant
 $J_{CP}$ that is derived from PMNS matrix elements as follows:
%$\delta_{CP}$ derived from PMNS matrix elements $U_{\alpha i}$ such that
\begin{equation}
J_{CP} = \text{Im} [U_{e1} U_{\mu 2} U_{e 2}^* U_{\mu 1}^*] = s_{23} c_{23} s_{12} c_{12} s_{13} c^2_{13} \sin \delta_{CP}.
\end{equation}
Majorana phases are estimated in terms of other invariants $I_1$ and $I_2$ as follows:
\begin{equation}
I_1 = \text{Im}[U^*_{e1} U_{e2}] = c_{12} s_{12} c_{13}^2 \sin \left( \frac{\alpha_{21}}{2} \right), \
I_2 = \text{Im}[U^*_{e1} U_{e3}] = c_{12} s_{13} c_{13} \sin \left( \frac{\alpha_{31}}{2}  - \delta_{CP} \right).
\end{equation}
In addition, the effective mass for the neutrinoless double beta decay is written by
\begin{align}
\langle m_{ee}\rangle=\kappa|\tilde D_{\nu_1} \cos^2\theta_{12} \cos^2\theta_{13}+\tilde D_{\nu_2} \sin^2\theta_{12} \cos^2\theta_{13}e^{i\alpha_{21}}+\tilde D_{\nu_3} \sin^2\theta_{13}e^{i(\alpha_{31}-2\delta_{CP})}|,
\end{align}
where its value could be measured by KamLAND-Zen in future~\cite{KamLAND-Zen:2016pfg}. 
We will adopt the neutrino experimental data at 3$\sigma$ interval~\cite{Esteban:2018azc,Nufit} as follows:
\begin{align}
&{\rm NO}: \Delta m^2_{\rm atm}=[2.432, 2.618]\times 10^{-3}\ {\rm eV}^2,\
\Delta m^2_{\rm sol}=[6.79, 8.01]\times 10^{-5}\ {\rm eV}^2,\\
&\sin^2\theta_{13}=[0.02046, 0.02440],\ 
\sin^2\theta_{23}=[0.427, 0.609],\ 
\sin^2\theta_{12}=[0.275, 0.350],\nn\\
%%%
&{\rm IO}: \Delta m^2_{\rm atm}=[2.416, 2.603]\times 10^{-3}\ {\rm eV}^2,\
\Delta m^2_{\rm sol}=[6.79, 8.01]\times 10^{-5}\ {\rm eV}^2,\\
&\sin^2\theta_{13}=[0.02066, 0.02461],\ 
\sin^2\theta_{23}=[0.430, 0.612],\ 
\sin^2\theta_{12}=[0.275, 0.350].\nn
\end{align}

\noindent
{\bf \underline{Non-unitarity}}: \\
Here, let us briefly discuss non-unitarity matrix $U'_{PMNS}$.
This is typically parametrized by the form 
\begin{align}
U'_{PMNS}\equiv \left(1-\frac12 FF^\dag\right) U_{PMNS},
\end{align}
where $F\equiv  (M^*)^{-1} m^T_D$ is a hermitian matrix, and $U'_{PMNS}$ represents the deviation from the unitarity. 
The global constraints are found via several experimental results such as the SM $W$ boson mass $M_W$, the effective Weinberg angle $\theta_W$, several ratios of $Z$ boson fermionic decays, invisible decay of $Z$, electroweak universality, measured Cabbibo-Kobayashi-Maskawa, and lepton flavor violations~\cite{Fernandez-Martinez:2016lgt}.
The result is then given by~\cite{Agostinho:2017wfs}
\begin{align}
|FF^\dag|\le  
\left[\begin{array}{ccc} 
2.5\times 10^{-3} & 2.4\times 10^{-5}  & 2.7\times 10^{-3}  \\
2.4\times 10^{-5}  & 4.0\times 10^{-4}  & 1.2\times 10^{-3}  \\
2.7\times 10^{-3}  & 1.2\times 10^{-3}  & 5.6\times 10^{-3} \\
 \end{array}\right].
\end{align} 
%%%
In our case, $F\equiv   (M^*)^{-1} m^T_D=\frac{v_1}{v_\varphi}(\tilde M^*)^{-1}\tilde m^T_D$.
Since we suppose to be $M >> m_D$ (coming from $v_\varphi>> v_1$) that is naturally realized by the difference  of breaking scale. Therefore, $v_\varphi$ is B-L breaking scale which is chosen to be higher than TeV scale, while $v_1$ is electroweak scale whose order is 0.1 TeV. Taking $v_\varphi \sim 10^2$ TeV we obtain $(v_1/v_\varphi)^2\approx 10^{-6}$, and we find
$|FF^\dag|\le 10^{-6}$ that is totally safe for the above bounds of the non-unitarity.
% even if $(\tilde M^*)^{-1}\tilde m^T_D \sim 1$.

 \section{Numerical analysis}
\label{sec:results}
In this section, we carry out numerical analysis searching for parameters satisfying neutrino data, 
and show our predictions.

%%%%%%%%%%%%%%%%%%%%%%%%%%%%%%%%%%%%%%%%%%%%%%%%%%%%%%%%%%%%%%%%%%%%%%%%%%%%%%%%%%%%
%%%%%%%%%%%%%%%%%%%
\begin{figure}[tb!]\begin{center}
\includegraphics[width=80mm]{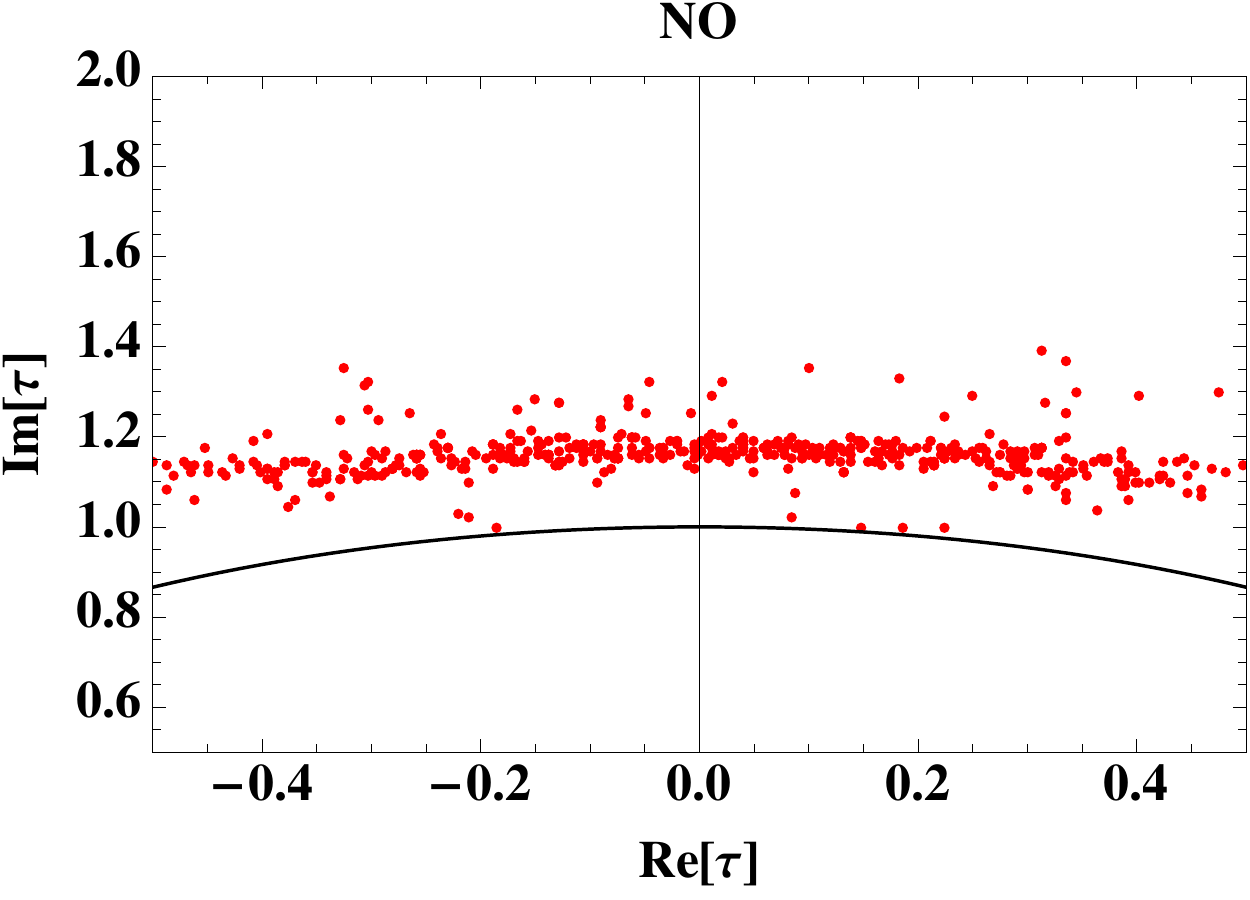} \ 
 \includegraphics[width=80mm]{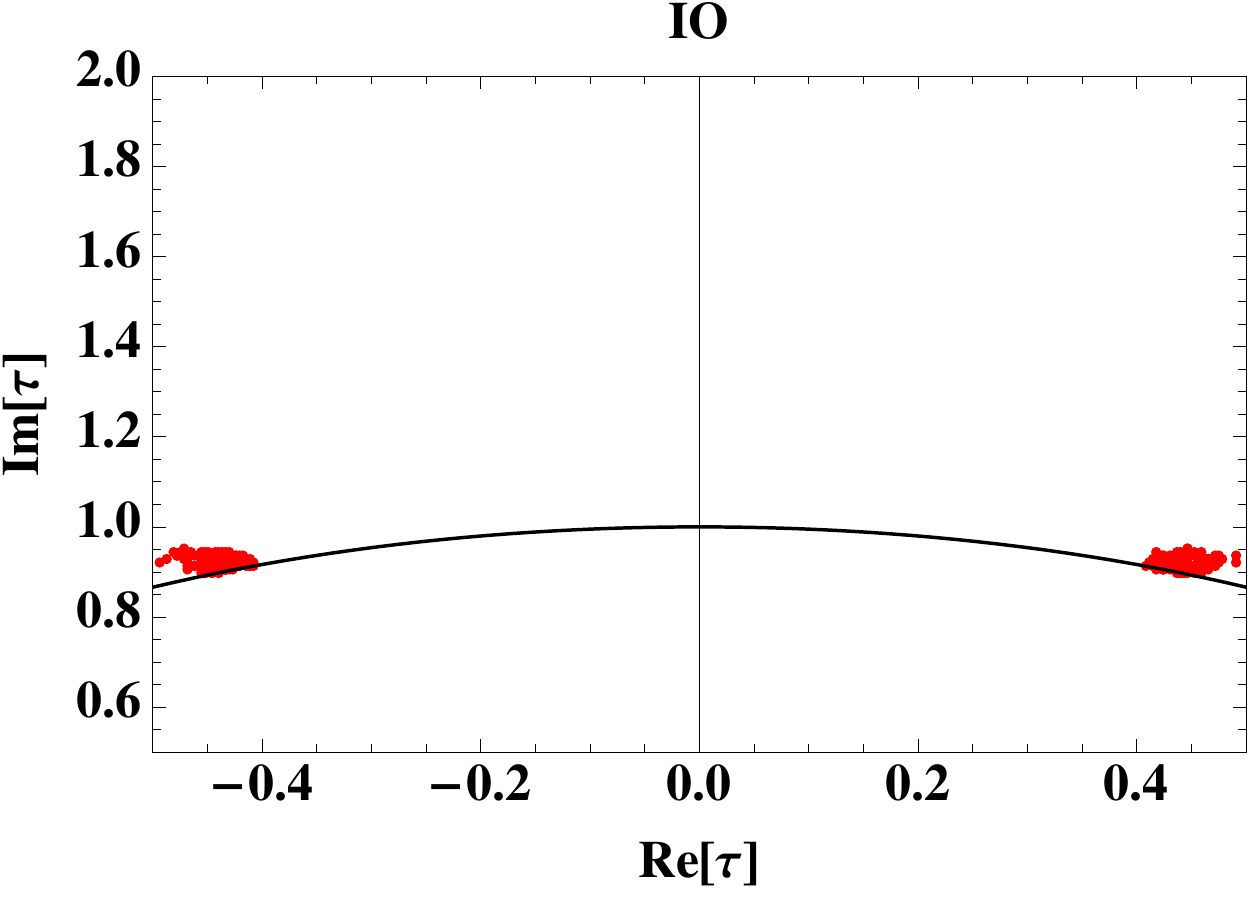}
\caption{The region of modulus $\tau$ satisfying neutrino data where the left and right panel correspond to NO and IO cases, respectively. The region above black solid curve correspond to fundamental domain of the modulus. }   
\label{fig:1}\end{center}\end{figure}
%%%%%%%%%%%%%%%%%%%
%%%%%%%%%%%%%%%%%%%%%%%%%%%%%%%%%%%%%%%%%%%%%%%%%%%%%%%%%%%%%%%%%%%%%%%%%%%%%%%%%%%%

%%%%%%%%%%%%%%%%%%%%%%%%%%%%%%%%%%%%%%%%%%%%%%%%%%%%%%%%%%%%%%%%%%%%%%%%%%%%%%%%%%%%
%%%%%%%%%%%%%%%%%%%
\begin{figure}[tb!]\begin{center}
\includegraphics[width=80mm]{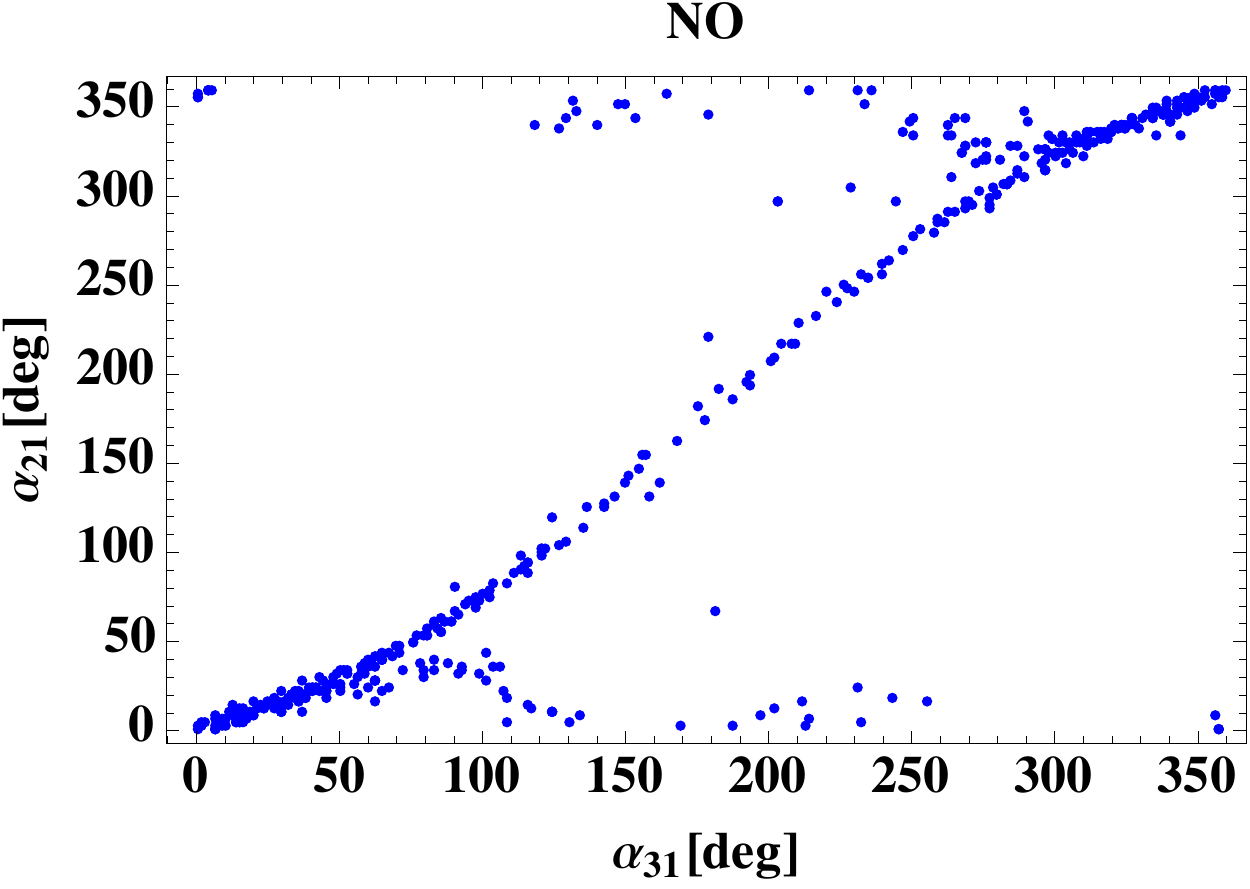} \ 
 \includegraphics[width=80mm]{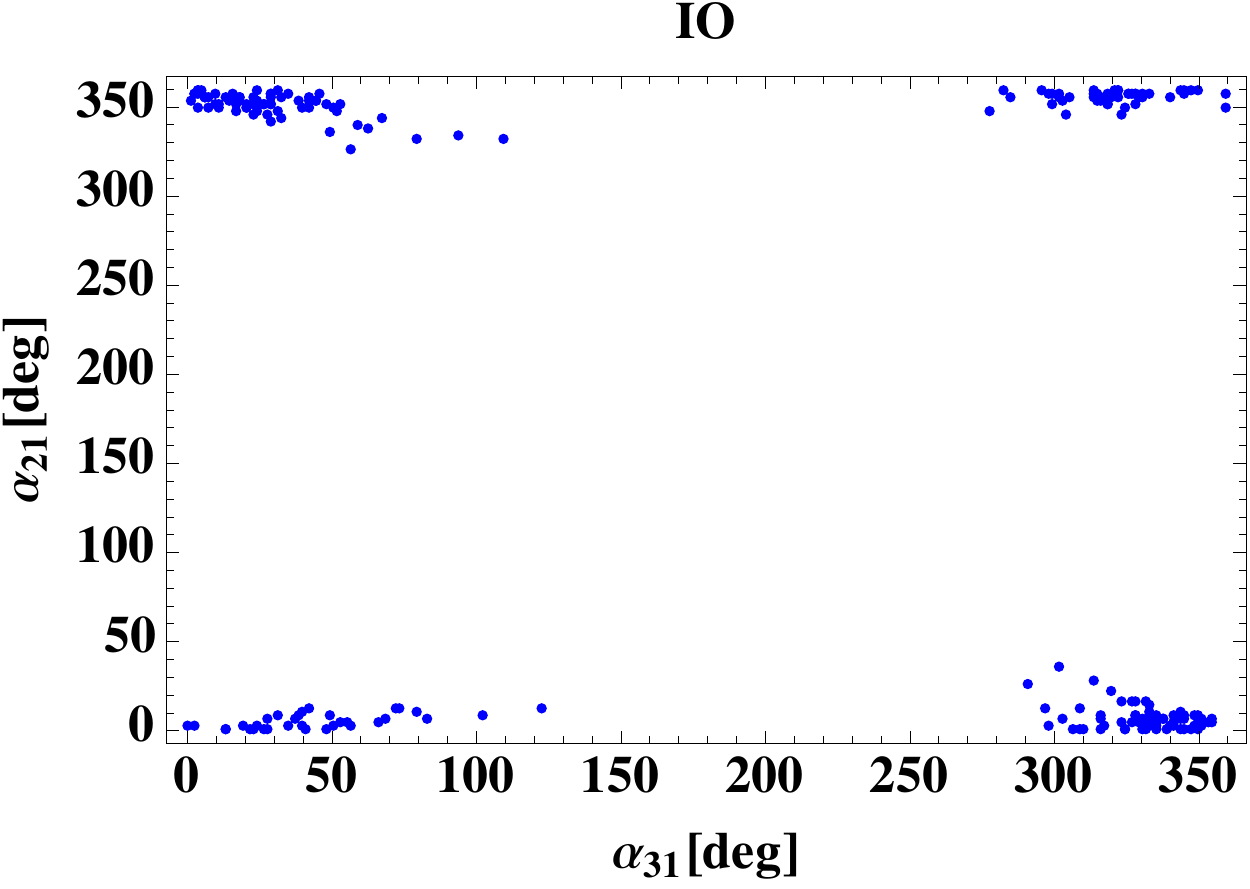}
  \includegraphics[width=80mm]{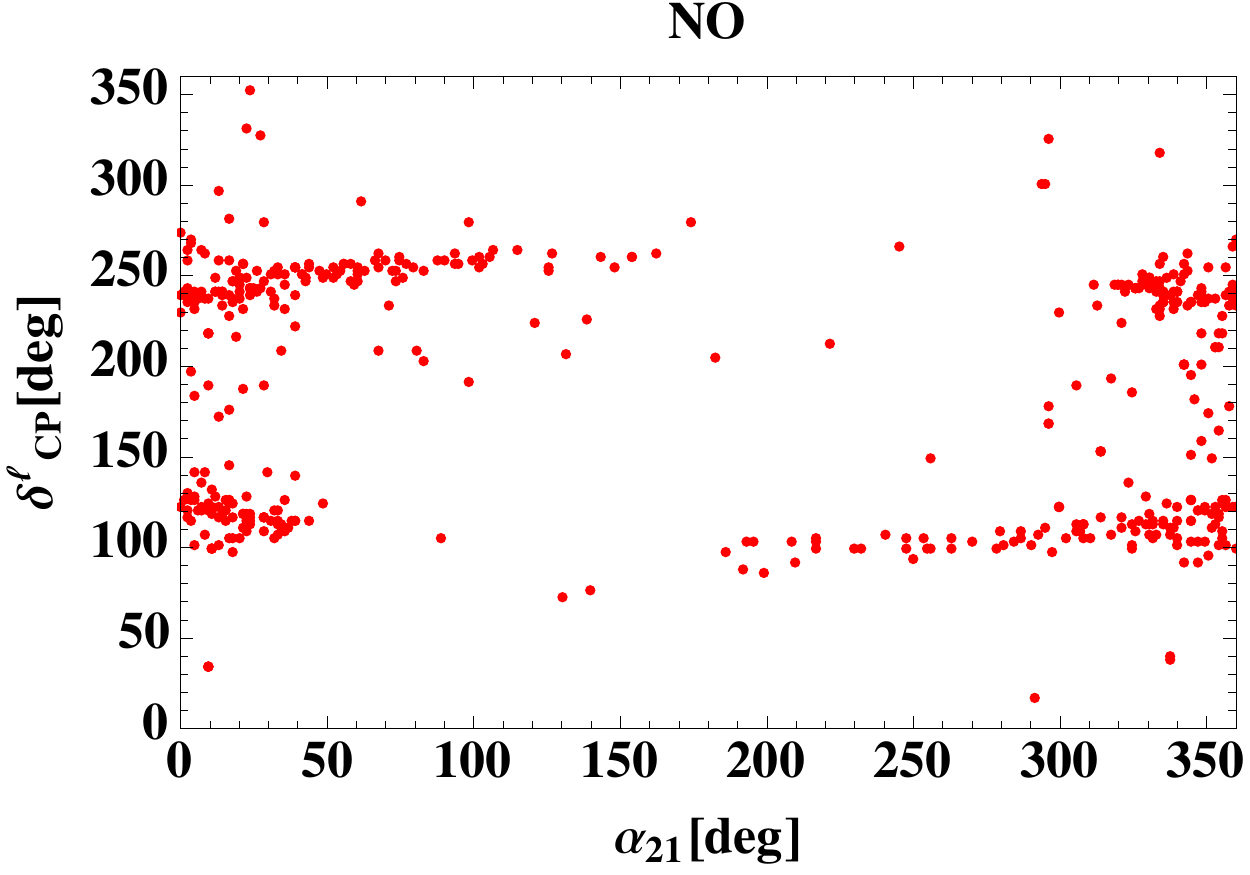} \ 
 \includegraphics[width=80mm]{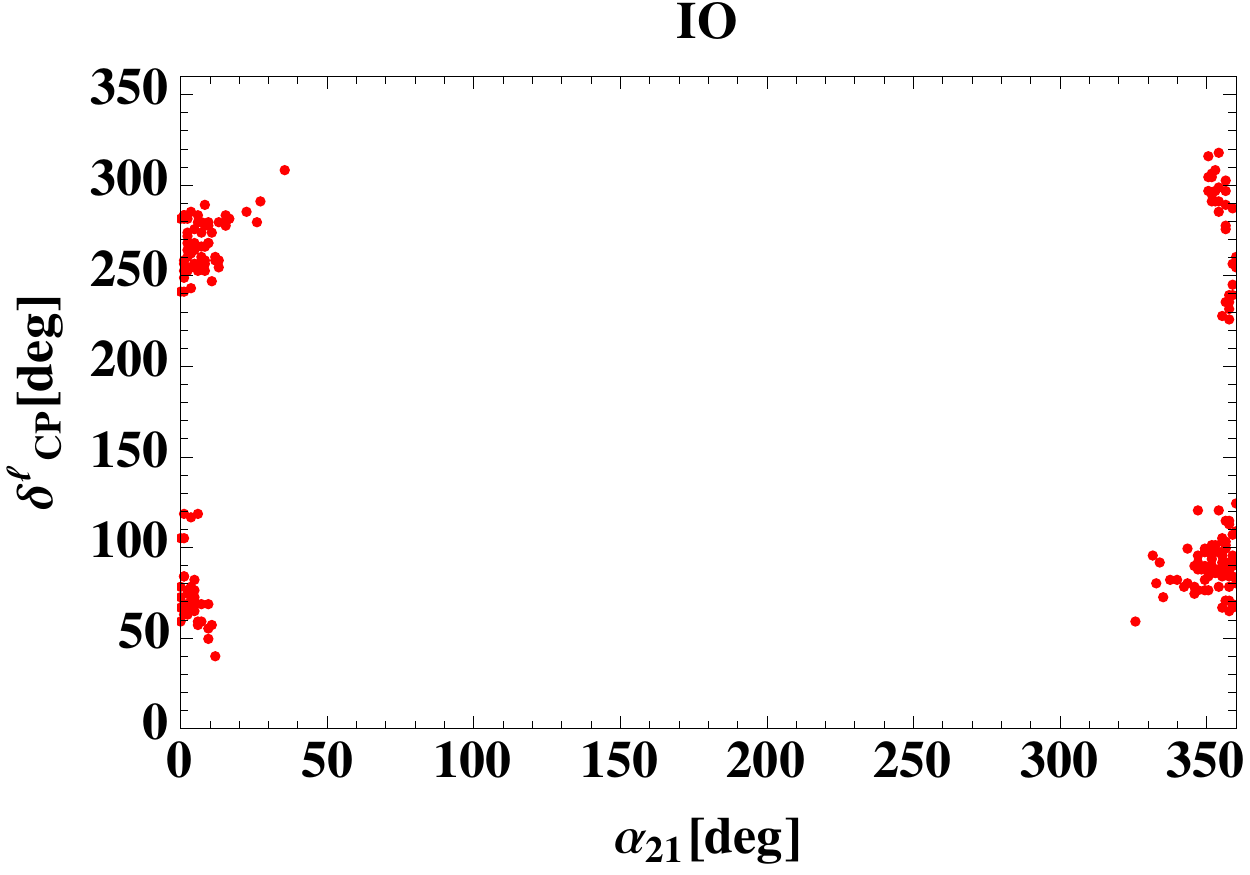}
\includegraphics[width=80mm]{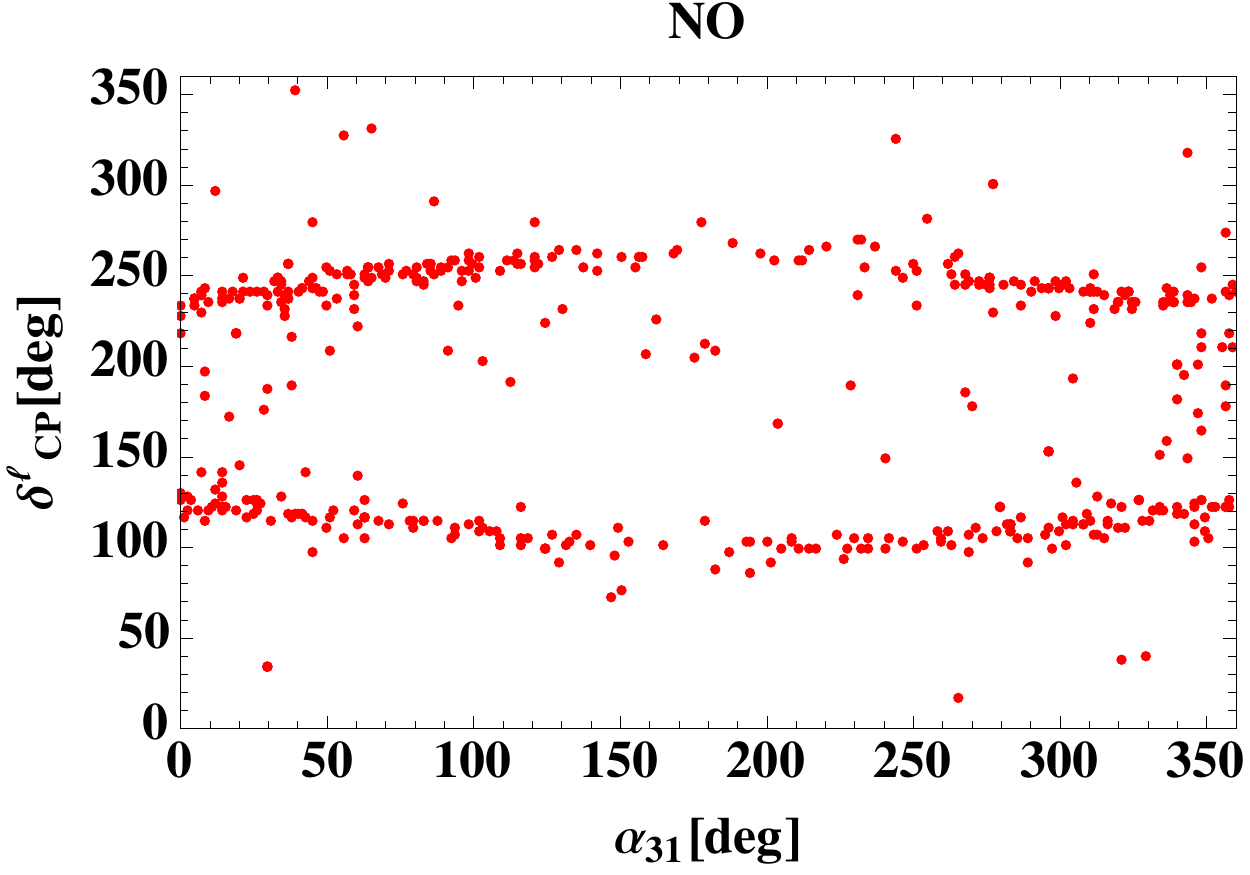} \ 
 \includegraphics[width=80mm]{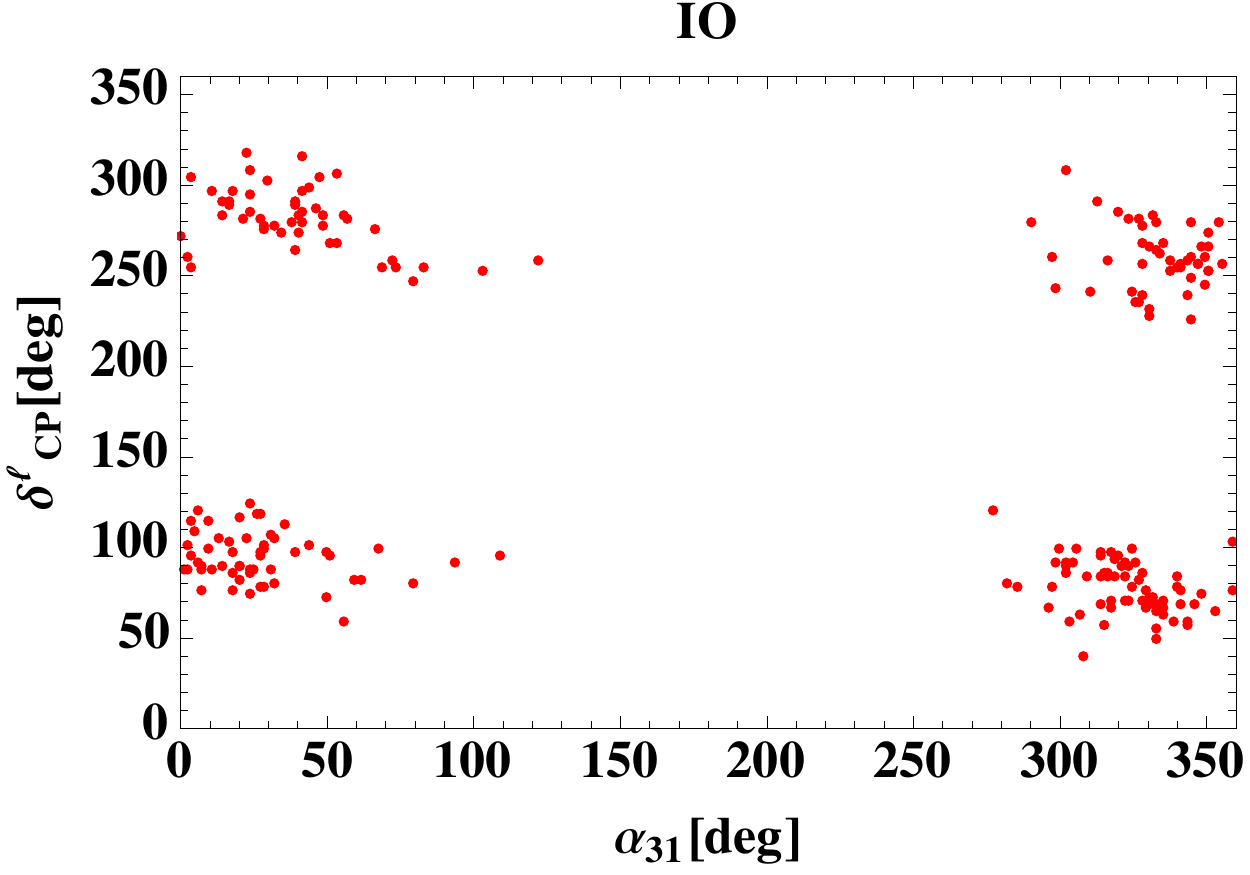}
\caption{Top figures: Predicted correlations between two Majorana phases $\alpha_{21}$ and $\alpha_{31}$.
Middle figures: Predicted correlations between $\alpha_{21}$ and Dirac CP phase $\delta^\ell_{\rm CP}$.
Bottom figures: Predicted correlations between $\alpha_{31}$ and $\delta^\ell_{\rm CP}$. 
The left-(right-)side figures correspond to NO(IO).}   
\label{fig:2}\end{center}\end{figure}
%%%%%%%%%%%%%%%%%%%
%%%%%%%%%%%%%%%%%%%%%%%%%%%%%%%%%%%%%%%%%%%%%%%%%%%%%%%%%%%%%%%%%%%%%%%%%%%%%%%%%%%%

In our numerical analysis, we scan free parameters in following ranges 
\begin{align}
&|{\rm Re}[\tau]| \in [0,0.5],\quad {\rm Im}[\tau]\in [0.5,2], \nonumber \\
& \{ |\alpha_{1}|,|\alpha_{2}|, |\gamma_1|, |\beta_1|, |\beta_2|, |\beta_3|, |\gamma_2|, |\gamma_3| \} \in [10^{-5},1.0], 
\end{align}
where couplings are taken to be complex values.
The parameters $\{a_\ell, b_\ell, c_\ell \}$ are fixed to reproduce the observed charged lepton masses where we numerically solve the conditions in Eq.~\ref{eq:l-cond}.

\noindent
{\bf \underline{Observable in neutrino sector}:}\\
As a result of numerical analysis, we find allowed parameter sets satisfying neutrino data for both NO and IO 
where we show the allowed region for modulus $\tau$ in Fig.~\ref{fig:1}.
%%%
Interestingly, we find a specific region at nearby a fixed point of $\tau=\omega$ in IO that is invariant under the $ST$ transformation, where $\omega\equiv e^{2\pi i/3}$. This can be considered as a remnant symmetry of $Z_3$ and is favored by a string theory~\cite{Kobayashi:2020uaj}. Thus, IO would be more interesting to be explored.

Fig.~\ref{fig:2} shows correlations among CP-violating phases where the left-side figures are for NO and the right-side ones are for IO.
Correlation between Majorana phases are given in the top plots. 
We find some correlation in NO while limited region are found in IO as $\{\alpha_{21}, \alpha_{31}\}$ being around $\{0^\circ -10^\circ, 0^\circ - 120^\circ \}$, $\{0^\circ -40^\circ, 290^\circ - 360^\circ \}$,
$\{330^\circ -360^\circ, 0^\circ - 120^\circ \}$ and $\{340^\circ -360^\circ, 280^\circ - 360^\circ \}$.
Correlations among Majorana phases and Dirac CP-phase are given in the middle and bottom panels.
We find that the  Dirac CP phase tends to be around $120^\circ$ and $240^\circ$ for NO, and $90^\circ$ and $270^\circ$ for IO.

%%%%%%%%%%%%%%%%%%%%%%%%%%%%%%%%%%%%%%%%%%%%%%%%%%%%%%%%%%%%%%%%%%%%%%%%%%%%%%%%%%%%
%%%%%%%%%%%%%%%%%%%
\begin{figure}[tb!]\begin{center}
\includegraphics[width=80mm]{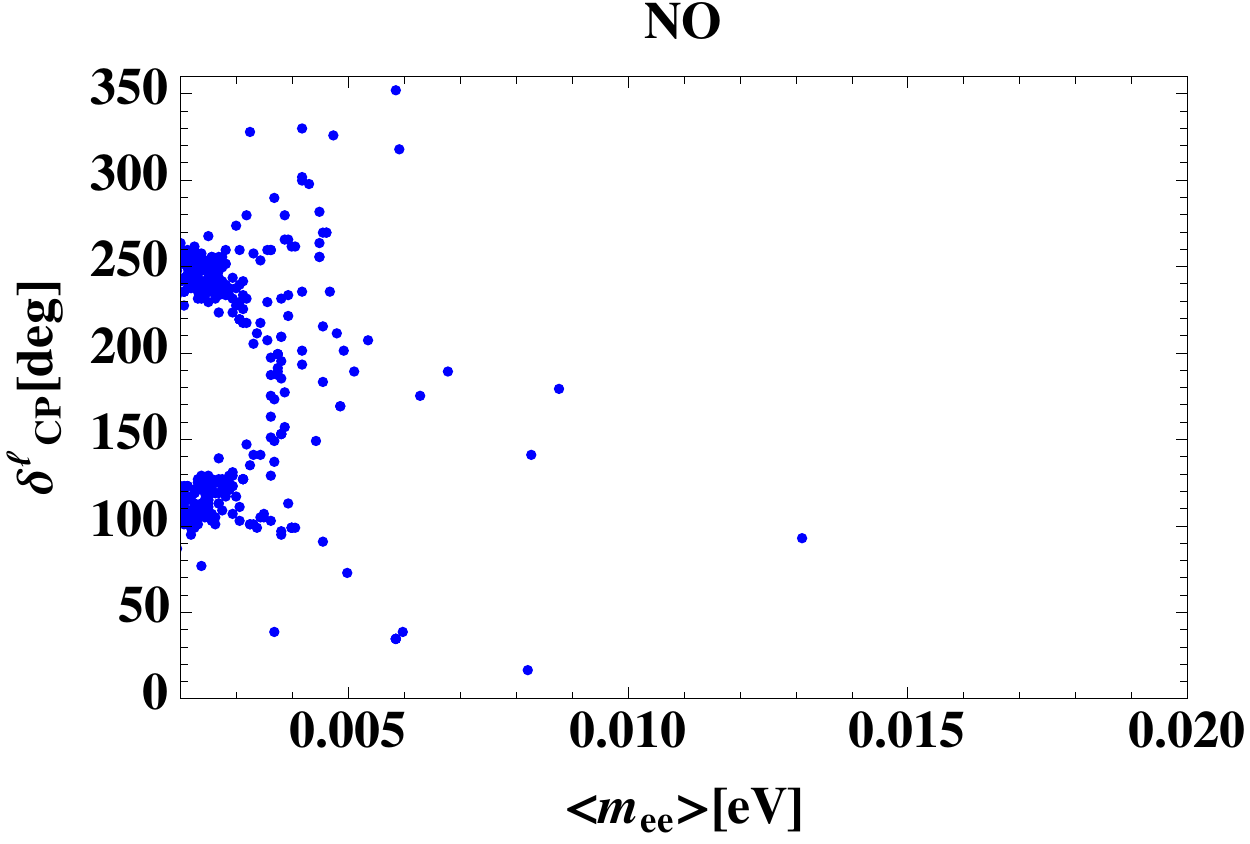} \ 
 \includegraphics[width=80mm]{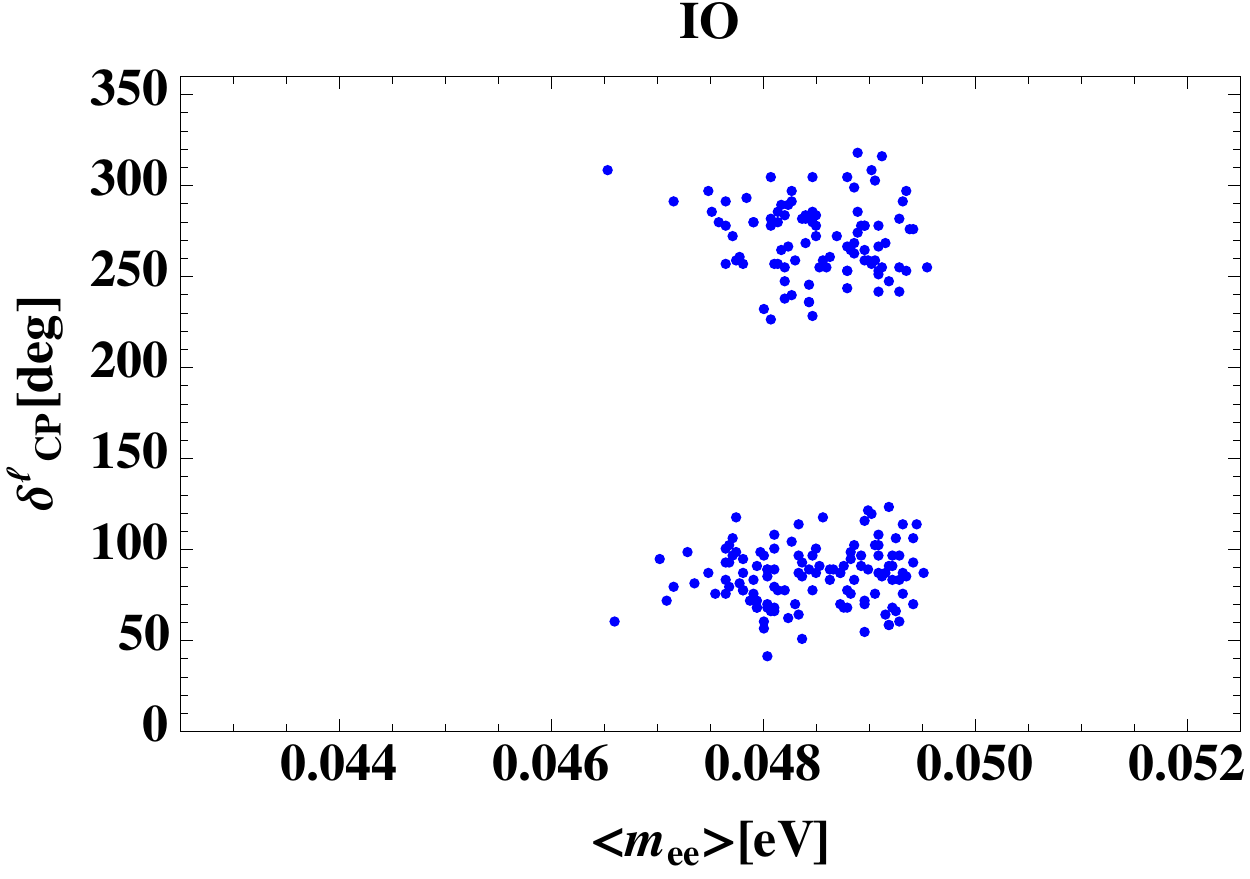}
 \includegraphics[width=80mm]{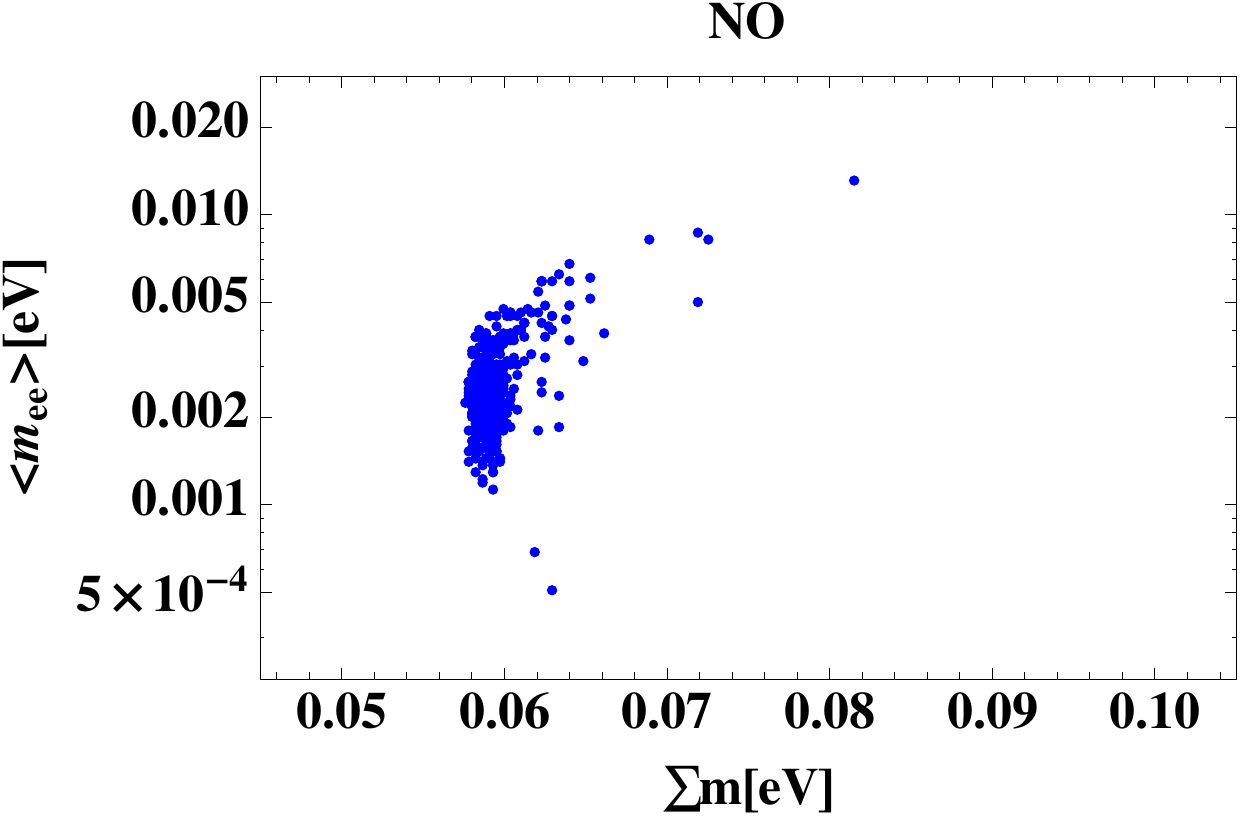} \ 
 \includegraphics[width=80mm]{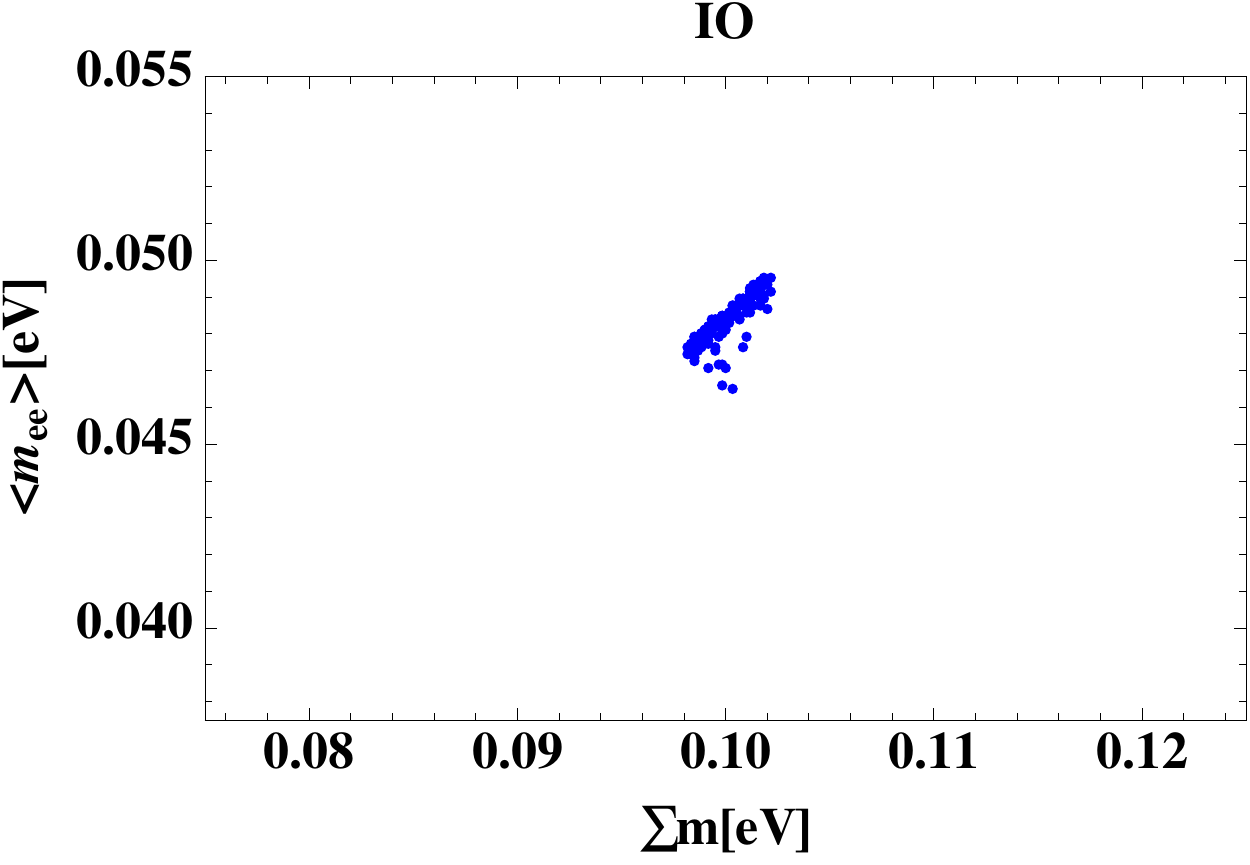}
\caption{Upper figures: Predicted correlation between the effective mass for the neutrinoless double beta decay $\langle m_{ee} \rangle$ and Dirac-CP phase $\delta^\ell_{CP}$. 
Lower figures: Predicted correlation between $\langle m_{ee} \rangle$ and sum of neutrino mass $\sum m$. The left-(right-)side figures correspond to NO(IO).}   
\label{fig:3}\end{center}\end{figure}
%%%%%%%%%%%%%%%%%%%
%%%%%%%%%%%%%%%%%%%%%%%%%%%%%%%%%%%%%%%%%%%%%%%%%%%%%%%%%%%%%%%%%%%%%%%%%%%%%%%%%%%%

The upper figures in Fig.~\ref{fig:3} show correlation between Dirac-CP phase and the effective mass for the neutrinoless double beta decay $\langle m_{ee}\rangle$, 
where the left-side one is for NO and the right-side one is for IO.
We find that $\langle m_{ee} \rangle$ tends to be small around  $\delta_{\rm CP}^\ell \sim 120^\circ$ and $240^\circ$ for NO 
while it is restricted around $0.046$ eV--$0.049$ eV for IO.
The lower figures in Fig.~\ref{fig:3} shows correlation between the sum of neutrino masses $\sum m(\equiv \kappa {\rm Tr}[\tilde D_\nu]$) versus $\langle m_{ee}\rangle$.
We have $0.058{\rm eV} \lesssim \sum m\lesssim0.082$eV for NO and $0.098{\rm eV} \lesssim \sum m\lesssim0.102$eV for IO.
Thus we find more restricted region for IO case.
In addition, both NO and IO satisfy cosmological constraint for sum of neutrino masses; $\sum m(\equiv$ Tr$[ D_\nu]) \lesssim 0.12$ eV.

%%%%%%%%%%%%%%%%%%%%%%%%%%%%%%%%%%%%%%%%%%%%%%%%%%%%%%%%%%%%%%%%%%%%%%%%%%%%%%%%%%%%
%%%%%%%%%%%%%%%%%%%
\begin{figure}[tb!]\begin{center}
\includegraphics[width=80mm]{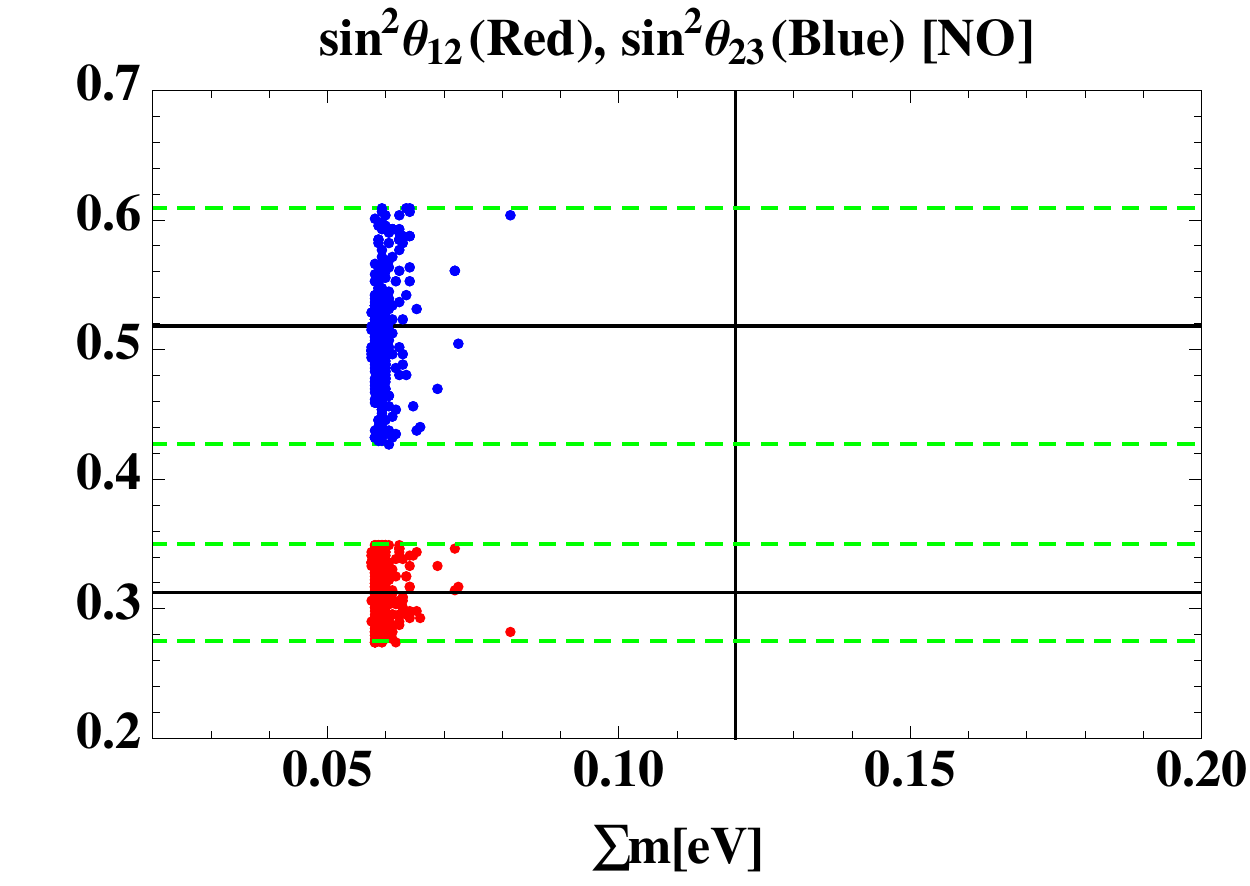} \ 
 \includegraphics[width=80mm]{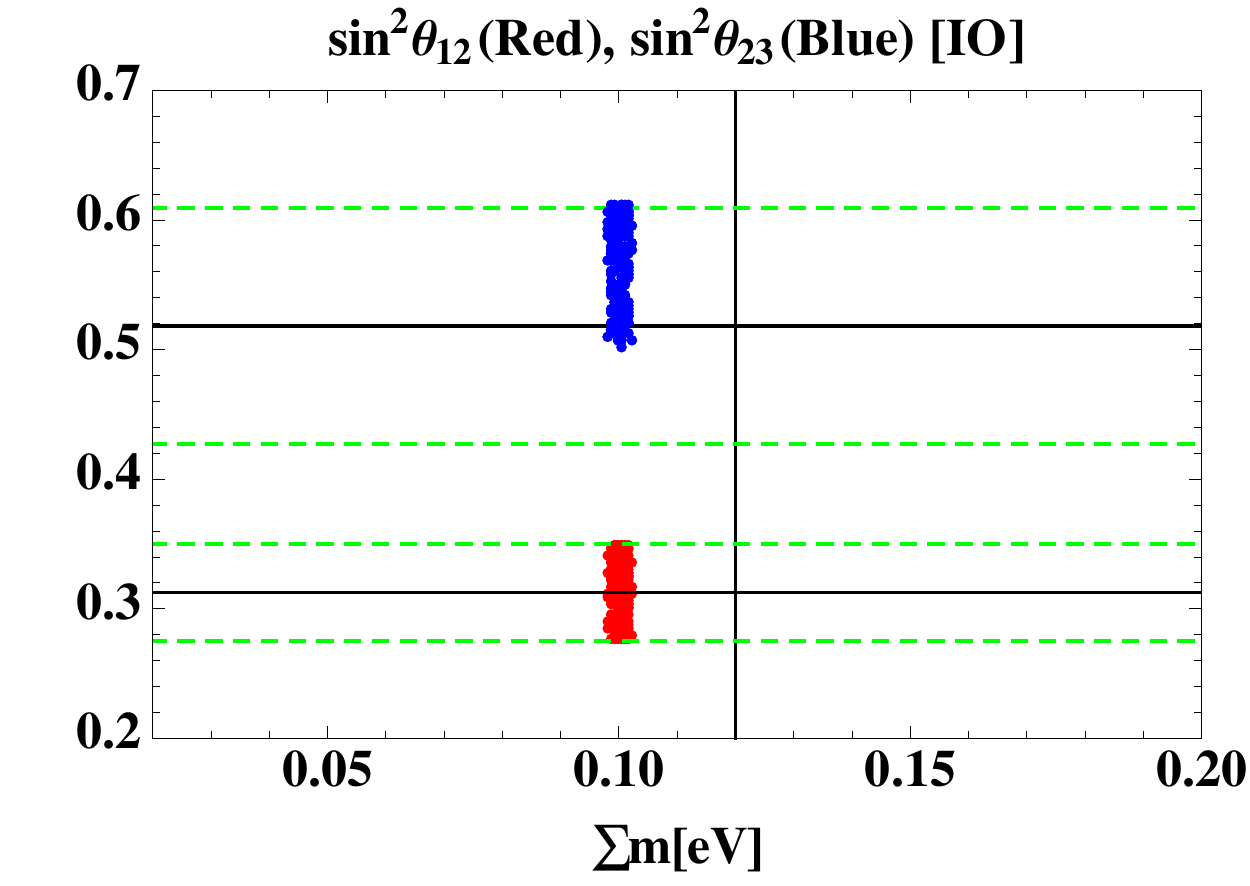}
 \includegraphics[width=80mm]{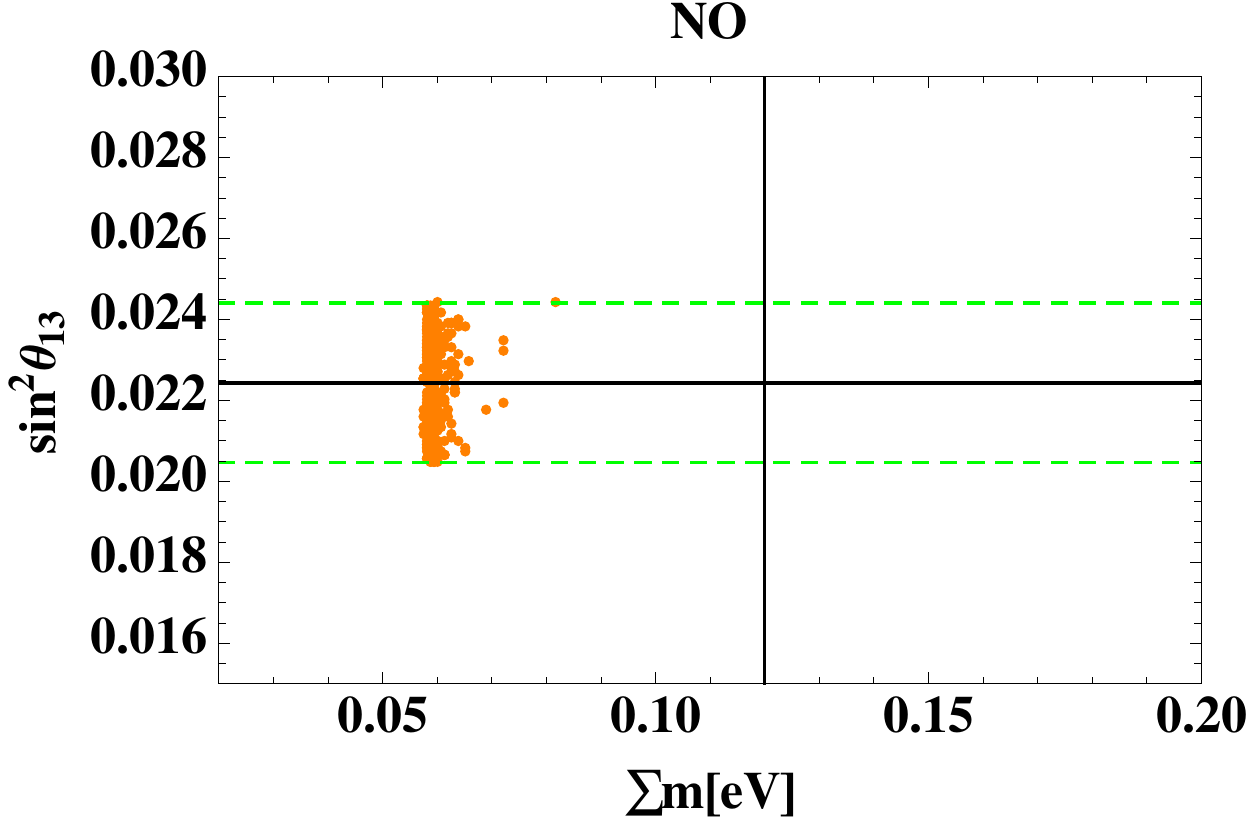} \ 
 \includegraphics[width=80mm]{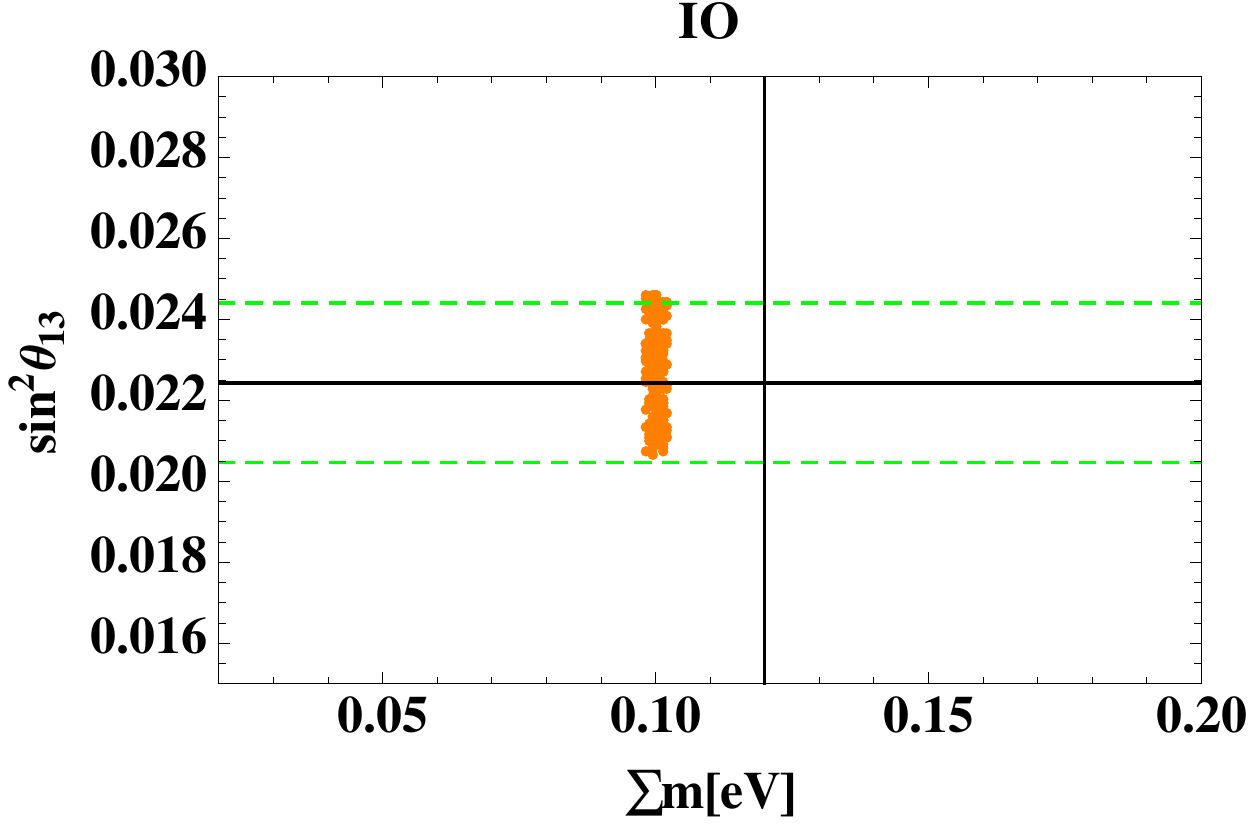}
\caption{The sum of neutrino masses $\sum m$ versus $\sin^2\theta_{12}$(red), $\sin^2\theta_{23}$(blue) for upper figures and $\sin^2\theta_{13}$ for lower figures,
where the left-(right-)side figure is NO(IO) and the vertical black line represents the cosmological constraint $\sum m_i\le0.12$ eV.}   
\label{fig:4}\end{center}\end{figure}
%%%%%%%%%%%%%%%%%%%
%%%%%%%%%%%%%%%%%%%%%%%%%%%%%%%%%%%%%%%%%%%%%%%%%%%%%%%%%%%%%%%%%%%%%%%%%%%%%%%%%%%%

Fig.~\ref{fig:4} shows relations between the sum of neutrino masses and $\sin^2\theta_{12} [\sin^2\theta_{23}]$ as indicated by red[blue] points for upper figures, 
and $\sin^2\theta_{13}$ for lower figures; the left-side figure is for NO and the right-side one is for IO.
The vertical black line indicates the cosmological constraint $\sum m \le0.12$ eV.
We find that the allowed region of $\sin^2\theta_{23}$ in case of IO favors the second octant region [0.5,0.623]
which could be more precisely measured by the future experiment~\cite{Srivastava:2017sno}.
On the other hand, the NO case has all allowed ranges of mixing angles.

%%%%%%%%%%%%%%%%%%%%%%%%%%%%%%%%%%%%%%%%%%%%%%%%%%%%%%%%%%%%%%%%%%%%%%%%%%%%%%%%%%%%
%%%%%%%%%%%%%%%%%%%
\begin{figure}[tb!]\begin{center}
\includegraphics[width=80mm]{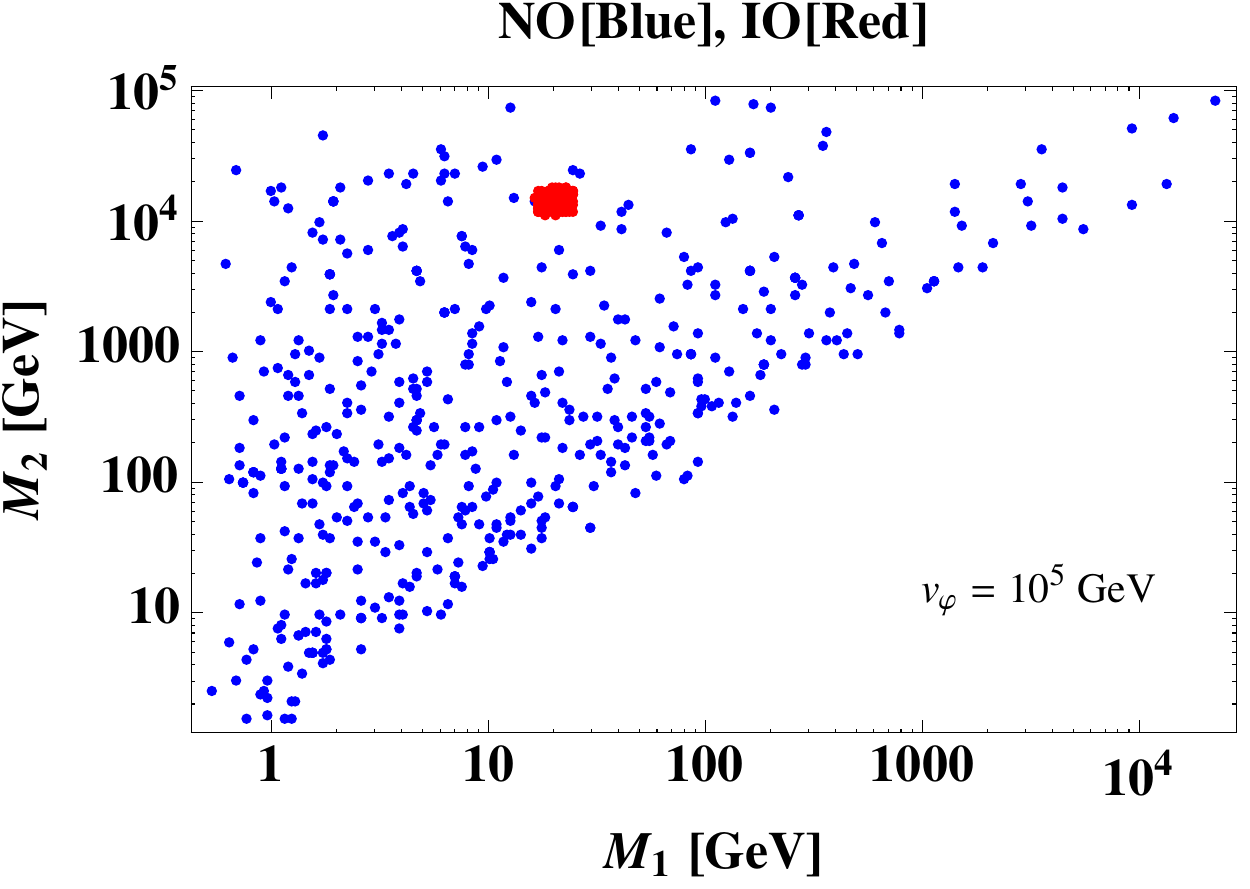} \ 
 \includegraphics[width=80mm]{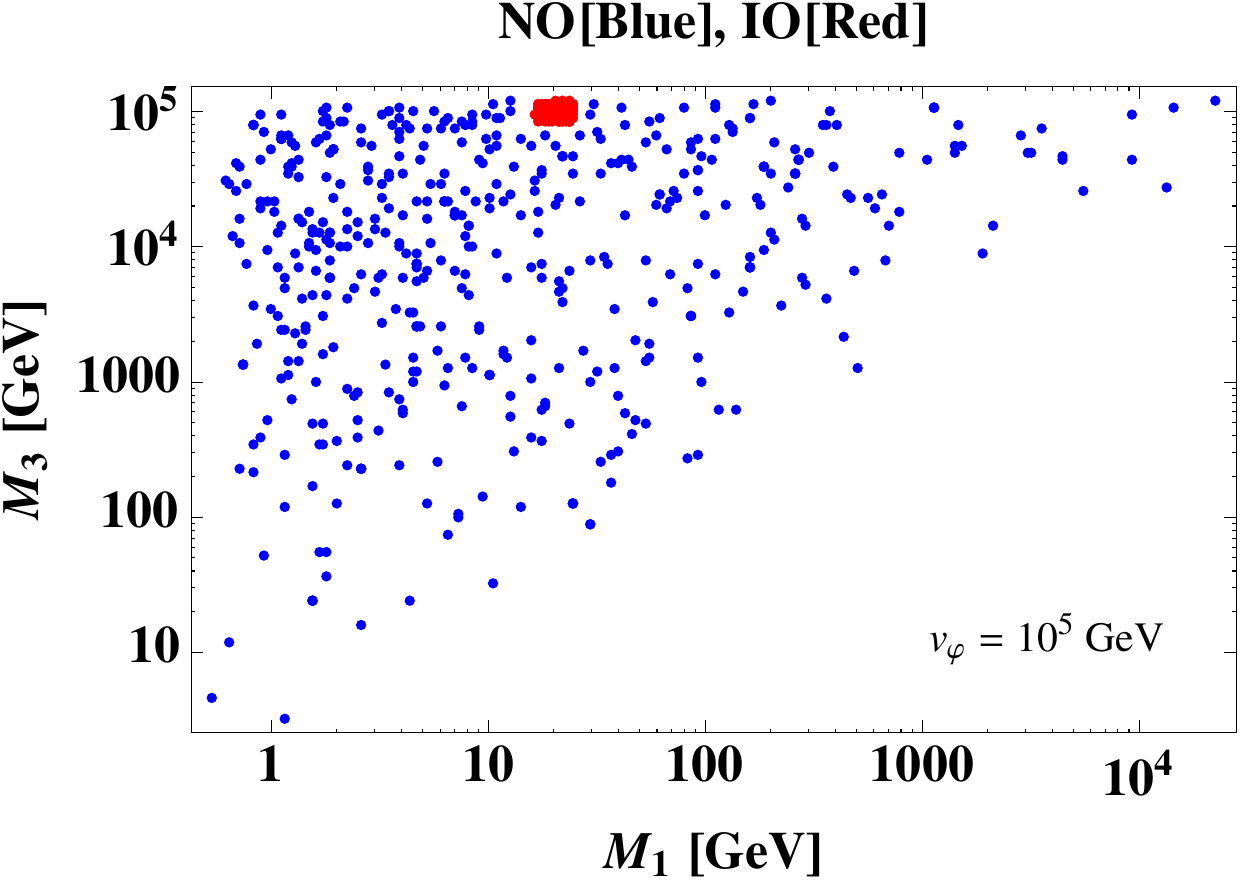}
\caption{Correlation among pseudo Dirac sterile neutrino masses where the blue points are for NO and the red ones are for IO.}   
\label{fig:5}\end{center}\end{figure}
%%%%%%%%%%%%%%%%%%%
%%%%%%%%%%%%%%%%%%%%%%%%%%%%%%%%%%%%%%%%%%%%%%%%%%%%%%%%%%%%%%%%%%%%%%%%%%%%%%%%%%%%

\noindent
{\bf \underline{Sterile neutrino mass}:}\\
In our model, sterile neutrinos are pseudo Dirac fermions whose masses are dominantly given by $M_{N_R S_L}$.
In Fig.~\ref{fig:5}, we also show the mass eigenvalues where we take $v_\varphi = 10^5$ GeV,  $M_{N_1} < M_{N_2} < M_{N_3}$ and blue(red) points corresponding to NO(IO). 
For NO, we do not find clear relation among the mass eigenvalues. On the other hand, we find limited preferred region of mass eigenvalues and hierarchy of $M_{N_1} \ll M_{N_{2,3}}$ in IO.
Some parameter regions could be tested at the LHC experiments, since sterile neutrino can be produced through $Z'$ boson.
In addition, $N_i$ can be produced through mixing with active neutrinos, $\theta_{N\nu} \simeq \sqrt{m_D M^{-1}}$, through the process $pp \to W \to \ell N_i$.
Detailed analysis of collider signature is beyond the scope of this work and will be given elsewhere.

\section{Summary and discussion }
\label{sec:conclusion}

We have constructed a linear seesaw model with local $U(1)_{B-L}$ and modular $A_4$ symmetry.
Majorana mass terms of sterile neutrinos are forbidden by $U(1)_{B-L}$ charge conservation and the nature of modular $A_4$ symmetry, 
and we can realize mass matrix for linear seesaw mechanism.
The Yukawa couplings for leptons are written by modular form which restricts the flavor structure of corresponding interactions. 

After formulating neutrino mass matrix, we have carried out numerical analysis searching for parameters satisfying neutrino data.
We have shown predicted observable such as Dirac CP phase, sum of neutrino masses, effective mass for neutrinoless double beta decay and sterile neutrino mass hierarchy.
Then some characteristic relations have been found for these observables.
In particular, 
%more restricted regions of observables are found for inverted ordering case. 
the case of IO favors a specific region at nearby $\tau=\omega$, which is favored by a string theory. Thus, our prediction would be very strong in this case.

%%%%%%%%%%%%%%%%%%%%%%%%%%%%%%%%%%%
\section*{Acknowledgments}
\vspace{0.5cm}
{\it
This research was supported by an appointment to the JRG Program at the APCTP through the Science and Technology Promotion Fund and Lottery Fund of the Korean Government. This was also supported by the Korean Local Governments - Gyeongsangbuk-do Province and Pohang City (H.O.). H. O. is sincerely grateful for the KIAS member.}
%%%%%%%%%%%%%%%%%%%%%%%%%%%%%%%%%%%
%%%%%%%%%%%%%%%%%%%%%%%%%%%%%%%%%%%
%%%%%%%%%%%%%%%%%%%%%%%%%%%%%%%%%%%

\section*{Appendix}

 %%%%%%%%%%%%%%%%%%%%%%%%%%%%%%%%%%%%%%%%%%%%%%%%%%%%%%%%%%%
Here we show some properties of $A_4$ modular symmetry framework. 
In general, the modular group $\bar\Gamma$ is the group of linear fractional transformation
$\gamma$ acting on the modulus $\tau$ 
which belongs to the upper-half complex plane and transforms as
\begin{equation}\label{eq:tau-SL2Z}
\tau \longrightarrow \gamma\tau= \frac{a\tau + b}{c \tau + d}\ ,~~
{\rm where}~~ a,b,c,d \in \mathbb{Z}~~ {\rm and }~~ ad-bc=1, 
~~ {\rm Im} [\tau]>0 ~.
\end{equation}
This is isomorphic to  $PSL(2,\mathbb{Z})=SL(2,\mathbb{Z})/\{I,-I\}$ transformation.
Then modular transformation is generated by two transformations $S$ and $T$ defined as follows; 
\begin{eqnarray}
S:\tau \longrightarrow -\frac{1}{\tau}\ , \qquad\qquad
T:\tau \longrightarrow \tau + 1\ ,
\end{eqnarray}
and they satisfy the following algebraic relations, 
\begin{equation}
S^2 =\mathbb{I}\ , \qquad (ST)^3 =\mathbb{I}\ .
\end{equation}

Here we introduce the series of groups $\Gamma(N)~ (N=1,2,3,\dots)$ which are defined by
 \begin{align}
 \begin{aligned}
 \Gamma(N)= \left \{ 
 \begin{pmatrix}
 a & b  \\
 c & d  
 \end{pmatrix} \in SL(2,\mathbb{Z})~ ,
 ~~
 \begin{pmatrix}
  a & b  \\
 c & d  
 \end{pmatrix} =
  \begin{pmatrix}
  1 & 0  \\
  0 & 1  
  \end{pmatrix} ~~({\rm mod} N) \right \}
 \end{aligned},
 \end{align}
and we define $\bar\Gamma(2)\equiv \Gamma(2)/\{I,-I\}$ for $N=2$.
Since the element $-I$ does not belong to $\Gamma(N)$
  for $N>2$ case, we have $\bar\Gamma(N)= \Gamma(N)$,
  that are infinite normal subgroup of $\bar \Gamma$ known as principal congruence subgroups.
   We thus obtain finite modular groups as the quotient groups defined by
   $\Gamma_N\equiv \bar \Gamma/\bar \Gamma(N)$.
For these finite groups $\Gamma_N$, $T^N=\mathbb{I}$  is imposed, and
the groups $\Gamma_N$ with $N=2,3,4,5$ are isomorphic to
$S_3$, $A_4$, $S_4$ and $A_5$, respectively \cite{deAdelhartToorop:2011re}.

Modular forms of level $N$ are 
holomorphic functions $f(\tau)$ which are transformed under the action of $\Gamma(N)$ given by
\begin{equation}
f(\gamma\tau)= (c\tau+d)^k f(\tau)~, ~~ \gamma \in \Gamma(N)~ ,
\end{equation}
where $k$ is the so-called as the  modular weight.

Here we discuss the modular symmetric theory framework without imposing supersymmetry explicitly, considering the $A_4$ ($N=3$) modular group. 
Under the modular transformation in Eq.(\ref{eq:tau-SL2Z}), a field $\phi^{(I)}$ is also transformed as 
\begin{equation}
\phi^{(I)} \to (c\tau+d)^{-k_I}\rho^{(I)}(\gamma)\phi^{(I)},
\end{equation}
where  $-k_I$ is the modular weight and $\rho^{(I)}(\gamma)$ denotes an unitary representation matrix of $\gamma\in\Gamma(2)$ ($A_4$ reperesantation).
Thus Lagrangian such as Yukawa terms can be invariant if sum of modular weight from fields and modular form in corresponding term is zero (also invariant under $A_4$ and gauge symmetry).

The kinetic terms and quadratic terms of scalar fields can be written by 
\begin{equation}
\sum_I\frac{|\partial_\mu\phi^{(I)}|^2}{(-i\tau+i\bar{\tau})^{k_I}} ~, \quad \sum_I\frac{|\phi^{(I)}|^2}{(-i\tau+i\bar{\tau})^{k_I}} ~,
\label{kinetic}
\end{equation}
which is invariant under the modular transformation and overall factor is eventually absorbed by a field redefinition consistently.
Therefore the Lagrangian associated with these terms should be invariant under the modular symmetry.

The basis of modular forms with weight 2, {\bf Y} = {$(y_{1},y_{2},y_{3})$},  transforming
as a triplet of $A_4$ is written in terms of the Dedekind eta-function  $\eta(\tau)$ and its derivative \cite{Feruglio:2017spp}:
%%%%%%%%%%%%%%%%%%%%%%%
\begin{eqnarray} 
\label{eq:Y-A4}
y_{1}(\tau) &=& \frac{i}{2\pi}\left( \frac{\eta'(\tau/3)}{\eta(\tau/3)}  +\frac{\eta'((\tau +1)/3)}{\eta((\tau+1)/3)}  
+\frac{\eta'((\tau +2)/3)}{\eta((\tau+2)/3)} - \frac{27\eta'(3\tau)}{\eta(3\tau)}  \right), \nonumber \\
y_{2}(\tau) &=& \frac{-i}{\pi}\left( \frac{\eta'(\tau/3)}{\eta(\tau/3)}  +\omega^2\frac{\eta'((\tau +1)/3)}{\eta((\tau+1)/3)}  
+\omega \frac{\eta'((\tau +2)/3)}{\eta((\tau+2)/3)}  \right) , \label{eq:Yi} \\ 
y_{3}(\tau) &=& \frac{-i}{\pi}\left( \frac{\eta'(\tau/3)}{\eta(\tau/3)}  +\omega\frac{\eta'((\tau +1)/3)}{\eta((\tau+1)/3)}  
+\omega^2 \frac{\eta'((\tau +2)/3)}{\eta((\tau+2)/3)}  \right)\,.
\nonumber
\end{eqnarray}
%%%%%%%%%%%%%%%%%%%%%
%
 Notice here that any singlet couplings under $A_4$ start from $-k=4$ constructed from the modular forms with $-k=2$ while it is absent if $-k=2$.

%\newref

\end{document}